\documentclass[useAMS,usenatbib]{mn2e}
\usepackage[british]{babel}
\usepackage{graphicx}
\usepackage{epstopdf}
\usepackage{color}
\usepackage{amsmath}
\usepackage{amssymb}

\title[IR spectroscopy of Cyg X--3]{Gemini/GNIRS infrared spectroscopy of the Wolf-Rayet stellar wind in Cygnus X--3}

\author[K. I. I. Koljonen]
{K.~I.~I.~Koljonen$^{1,2,3}$\thanks{email: karri.koljonen@utu.fi}, T.~J.~Maccarone$^{4}$
\\
$^{1}$Finnish Centre for Astronomy with ESO (FINCA), University of Turku, V\"ais\"al\"antie 20, 21500 Piikki\"o, Finland \\
$^{2}$Aalto University Mets\"ahovi Radio Observatory, PO Box 13000, FI-00076 Aalto, Finland \\
$^{3}$New York University Abu Dhabi, PO Box 129188, Abu Dhabi, UAE \\
$^{4}$Department of Physics and Astronomy, Texas Tech University, Box 41051, Lubbock, TX 79409-1051, USA}

\begin{document}

\pagerange{\pageref{firstpage}--\pageref{lastpage}}
\pubyear{2014}

\maketitle

\label{firstpage}

\begin{abstract} 
The microquasar Cygnus X--3 was observed several times with the Gemini North Infrared Spectrograph while the source was in the hard X-ray state. We describe the observed 1.0--2.4 $\mu$m spectra as arising from the stellar wind of the companion star and suggest its classification as a WN 4--6 Wolf-Rayet star. We attribute the orbital variations of the emission line profiles to the variations in the ionization structure of the stellar wind caused by the intense X-ray emission from the compact object. The strong variability observed in the line profiles will affect to the mass function determination. We are unable to reproduce earlier results, from which the mass function for the Wolf-Rayet star was derived. Instead, we suggest that the system parameters are difficult to obtain from the infrared spectra. We find that the near-infrared continuum and the line spectra can be represented with non-LTE Wolf-Rayet atmosphere models if taking into account the effects arising from the peculiar ionization structure of the stellar wind in an approximative manner. From the representative models we infer the properties of the Wolf-Rayet star, and discuss possible mass ranges for the binary components.  
\end{abstract}


\section{Introduction} \label{introduction}

One of the most peculiar sources amongst microquasars is Cygnus X-3 (Cyg X--3). It is known for massive outbursts that emit throughout the electromagnetic spectrum from radio to $\gamma$-rays and produce major radio flaring episodes usually with multiple flares that peak up to 20 Jy (e.g., \citealt{waltman95}), making it the single most radio luminous object in our Galaxy at its peak. The most striking feature of its X-ray lightcurve \citep{parsignault72} is a strong 4.8-hour periodicity which is attributed to the orbital modulation of the binary. The same periodicity has been observed in the infrared, that is in phase with the X-rays but has only 10\% modulation while the X-rays are modulated by 50\% \citep{becklin73,becklin74,mason76,mason86}. 

The distance to Cyg X--3 has been recently estimated to be 7.4 kpc \citep{mccollough16}. Since Cyg X--3 lies in the Galactic plane, interstellar extinction has prevented the detection of the optical/UV counterpart, thus rendering the usual identification techniques of the stellar companion obsolete, and promoting the use of infrared observations. The infrared spectrum of Cyg X--3 resembles most closely a Wolf-Rayet (WR) star \citep{vankerkwijk92,vankerkwijk93,vankerkwijk96}. 

Whether the compact object in Cyg X--3 is a black hole or a neutron star is not certain. \citet{zdziarski13} favour a low-mass black hole based on orbital kinematics measured using infrared and X-ray emission lines \citep{hanson00,vilhu09}. On the other hand, based on X-ray spectral modelling, the mass of the black hole could be as large as 30 $M_{\odot}$ \citep{hjalmarsdotter08}. Also, a neutron star as its compact object has never decisively been ruled out. The mass estimate in \citet{zdziarski13} relies in part to radial velocity measurements of certain helium and nitrogen lines \citep{hanson00} using quiescent K-band spectra from the Multiple Mirror Telescope \citep{fender99}. Although these observations give a nearly constant value for the radial velocity amplitude of these lines, the systemic velocity changes from line to line by as much as 270 km s$^{-1}$. The reason for this could be the turbulent motion of the WR stellar wind. 

In \citet{hanson00}, the mass function of the WR star was derived from the radial velocity curve of what was identified as a He I 2p--2s absorption line, motivated by its proper phasing and assuming that the line was tracing the stellar wind motion in such a way that it traces the binary motion of the WR star. On the other hand, it is not clear whether the infrared line velocities have anything to do with the orbital motions of the binary, but rather reflect the velocity field of the stellar wind. In addition, the observations were taken during an outburst in Cyg X--3, when a strong, double-peaked He I 2p--2s emission line was present complicating the line region. Since 1999, no similar high-resolution infrared observations have been published, and this issue remains to be solved. 

Cyg X--3 is extremely compact system as implied by its orbital period. Regardless of the nature of the compact object, if the WR companion star is correctly identified, it introduces a massive stellar wind component to the system that is likely to figure prominently in much of the phenomenology of the system. The stellar wind is embedded in a region of high X-ray energy density. This suggests that the luminous X-ray source should have a strong influence on the behaviour of the stellar wind. In addition to providing the fuel for accretion of matter to the compact object, the stellar wind has been proposed to be a major component of the infrared emission \citep{vankerkwijk92,vankerkwijk93,vankerkwijk96}, modifying the hard X-ray emission by Compton down-scattering \citep{zdziarski10}, modifying the soft X-ray emission by absorption and re-emission \citep{szostek08,zdziarski10}, producing photoionized X-ray emission \citep{paerels00}, and emitting seed photons to be scattered by the relativistic electrons in the jet to produce $\gamma$-rays \citep{tavani09,fermi09,dubus10}.

The infrared spectrum shows a wealth of helium and nitrogen lines, whose radial velocity changes with the phase of the binary. However, distinct from a normal binary with blueshift and redshift occurring at the ascending and descending node, the maximum blueshift is observed at X-ray phase $\phi$ = 0.0, and maximum redshift at X-ray phase $\phi$ = 0.5. This has been previously explained with a model where the lines are formed in the X-ray shadow, behind the WR star, where the intense X-ray emission from the compact object does not ionize the gas, and where a normal line-driven wind can be formed \citep{vankerkwijk96}. Similarly, the relative phasing of the infrared and X-ray continua can be understood so that the cool part of the wind is more opaque, thus shadowing the hot part and resulting in infrared minimum at X-ray phase $\phi$ = 0.0 \citep{vankerkwijk93}.     

Due to these observables Cyg X--3 is by definition a unique source amongst high-mass X-ray binaries since it harbors an atypical companion star and has a very short orbital period. On the premise that the binary constitutes a WR companion and a black hole, this uniqueness has been established also through population studies \citep{lommen05}. Similar systems have been observed in other nearby galaxies: IC 10 X--1 \citep{prestwich07,silverman08}, NGC 300 X--1 \citep{carpano07,crowther10,binder11}, and WR/black hole candidates in NGC 4490 \citep{esposito13} and NGC 253 \citep{maccarone14}. The combination of short orbital period, and the presence of a compact object with a massive companion star makes it a good prototype candidate for being a progenitor of a gravitational wave source \citep{belczynski13}.

In this paper we use Gemini Near-IR Spectrograph (GNIRS) on the 8.1 m Gemini North telescope to measure the infrared ($H$-, $J$- and $K$-band) spectrum of Cyg X--3 to study the stellar wind component in detail. In Section \ref{observations}, we present the observations and the process to obtain the reduced infrared spectra. In Section \ref{results}, we present the properties of the spectra including the continuum and line spectra, and show that they are best represented by a WN 4--6 star. We compare the averaged line spectrum to the non-LTE atmosphere models of WR stars, and show that representative models to the data can be found by taking into account the impact of the ionizing X-ray radiation to the stellar wind. In addition, we study the line profile through the orbital phase, present radial velocity profile and full width at half maximum (FWHM) of the emission lines and one absorption line, and compare them to the previous results from literature. In Section \ref{discussion}, we discuss that the results indicate a complicated ionization structure in the wind, or possibly a presence of a shock between the compact object and stellar wind, and likely modifications contributed by clumps in the wind. In addition, based on the modelling of the line spectrum, we derive estimates for the stellar wind parameters, and discuss probable mass ranges for the binary components. We present our conclusions in Section \ref{conclusions}.          

\section{Observations and data reduction} \label{observations}

The infrared spectra were acquired as a poor weather program with cloudy or poor seeing conditions. Some datasets were omitted from the analysis due to low statistical significance resulting from very poor conditions. We used the cross-dispersed mode of the GNIRS, ``short blue'' camera, 32 l/mm grating and 0.3'' wide/7'' long slit. This mode gives simultaneous spectral coverage from $\sim$1.0--2.4 $\mu$m with resolving power of R$\sim$1400 and a pixel scale of 0.15''/pixel. The telescope was nodded along the slit by $\sim$3'' in ABBA-type sequence, where the individual exposure time for one ABBA-sequence was $\sim$10 minutes. The total exposure times varied depending on the queue and weather conditions, typically being $\sim$1 hour long, and Cyg X--3 was successfully observed during six nights in June, July and November 2015. The list of the observations used can be found in Table A1.    

The raw spectra were reduced using {\sc XDGNIRS} pipeline version 2.2.6.\footnote{The code can be accessed at the Gemini Data Reduction User Forum: http://drforum.gemini.edu/forums/gemini-data-reduction/} designed to reduce GNIRS cross-dispersed spectrum using mainly the data reduction tasks of the Gemini IRAF package. {\sc XDGNIRS} produces a roughly flux-calibrated spectrum from an ABBA-set of raw science and calibration files. In the following, the reduction steps are briefly introduced. First, the data are cleaned of any patterned noise and radiation events are removed. The source and standard star files are then flat-fielded, sky-subtracted, rectified and wavelength-calibrated. After combining the ABBA-set of files, the spectrum of each spectral order is then extracted using an aperture of 12 pixels (1.8'') along the slit. For telluric absorption line removal, we used type A1V standard star (HIP 99893 or HIP 103108) observed before or after each observation run. The intrinsic absorption lines are removed from the standard star spectrum by fitting Lorentz profiles to the lines, after which the source spectrum is corrected by the standard star spectrum. In some regions (1.34--1.45 $\mu$m and 1.80--1.95 $\mu$m) the atmosphere is dominant and these are cut from the final spectra. To produce roughly flux-calibrated spectra, each order is multiplied by a blackbody spectrum of appropriate temperature scaled to the $K$-band magnitude of the standard star. Finally, the orders are combined and the inter-order offsets are removed to produce the final spectrum. At the end of reduction process we went through the spectrum and looked for any obvious, spurious artefacts and interpolated over them if necessary. 

\section{Results} \label{results}

\subsection{The 1.0--2.4 $\mu$m spectrum} \label{spectrum}

\begin{figure}
\begin{center}
\includegraphics[width=0.5\textwidth]{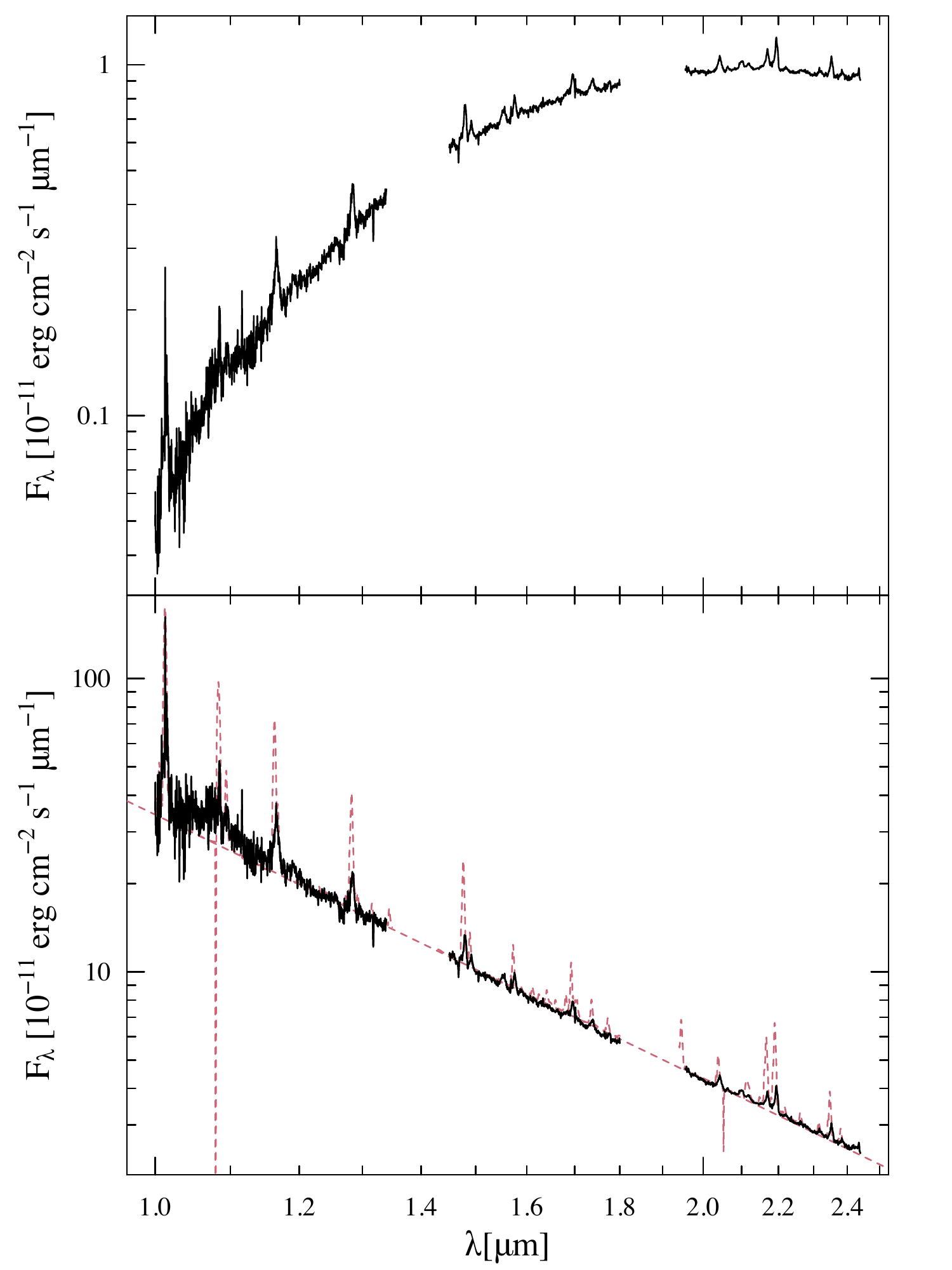}
\end{center}
\vspace{-12pt}
\caption{The 1.0--2.4 $\mu$m Cyg X--3 spectrum as measured on 16 June 2015 taken during X-ray phase $\phi$ = 0.5 with reddening (upper panel) and de-reddened (lower panel) using Stead \& Hoare extinction law with A$_{K}$=1.44. The dashed line in the lower panel is the spectral energy distribution ($F_{\lambda} \propto \lambda^{\beta}$) from one of the representative non-LTE atmosphere models (WNL 08-13; see Section \ref{wrmodel}) with an index $\beta=-3.0$. The obvious discrepancy in the emission line fluxes is discussed in detail in Section \ref{wrmodel}} \label{spec}
\end{figure}

The stellar type identified with the Cyg X--3 infrared spectrum is that of a WR star \citep{vankerkwijk92,vankerkwijk96}. The infrared bands previously used have been $K$- and $H$-band \citep{fender99} and $K$- and $I$-band \citep{vankerkwijk92,vankerkwijk93,vankerkwijk96}. We observed Cyg X--3 in the $J$-band in addition to the $K$- and $H$-band. The broadband spectrum is essential to make the comparison to a WR atmosphere model spectrum more robust, and to distinguish between different stellar types. Fig. \ref{spec} shows one 10 min exposure spectrum on 16 June 2015 taken during X-ray phase $\phi$ = 0.5, i.e. the infrared maximum, displaying heavy extinction and a wealth of emission lines. We have excluded regions where the atmospheric emissions dominate.

Assuming that the infrared lines arise from the WR wind, the infrared continuum should resemble the continuum from the WR star as well, namely a power law in the form $F_{\lambda} \propto \lambda^{\beta}$. In a global study by \citet{morris93}, WR stars were found to have $<\beta> = -2.85 \pm 0.4$. In the case of Cyg X--3, this approximately corresponds to A$_{K}$=1.4$\pm$0.1 when using the \citet{stead09} extinction law, where the near-IR extinction is a power law with $A_{\lambda} \propto \lambda^{-2.14}$. The K-band extinction is much lower than the previous value derived in \citealt{vankerkwijk96}, A$_{K}$=2.1$\pm$0.4, due to more modern value for the the near-IR extinction law (as compared to $A_{\lambda} \propto \lambda^{-1.7}$; \citealt{mathis90}), and will affect to the classification of the WR star and its distance estimate. 

We estimate the K$_{S}$-band magnitude 11.5 mag from the average flux density. This is in line with the 2MASS value of 11.6 mag, and value derived in \citet{mccollough10}: 11.4--11.7 mag, which was based on a multi-year monitoring data from PAIRITEL (The Peters Automated Infrared Imaging Telescope; \citealt{bloom06}). The distance to Cyg X--3 is somewhat uncertain. \citet{mccollough16} derive two probable distances of 7.4$\pm$1.1 kpc or 10.2$\pm$1.2 kpc, the former being slightly preferred (62\% probability). Using these inferred distances, A$_{K}$, and the apparent K-band magnitude range from above, the absolute K-band magnitude is then $M_{K_{S}}$ = $-$4.2$\pm$0.4 or $M_{K_{S}}$ = $-$4.9$\pm$0.3, respectively. Depending on the distance, this range indicates a weak-lined WN5, or weak-lined WN6/broad-lined WN4--5 as the WR sub-type \citep{rosslowe15}. However, as considerable scatter exists in the absolute magnitude estimates, as well as in the calibration values of WN spectral types in \citet{rosslowe15}, the classification based on the absolute K-band magnitude is not robust (more detailed estimate follows in Section \ref{wrtype}). 

\begin{figure*}
\begin{center}
\includegraphics[width=0.97\textwidth]{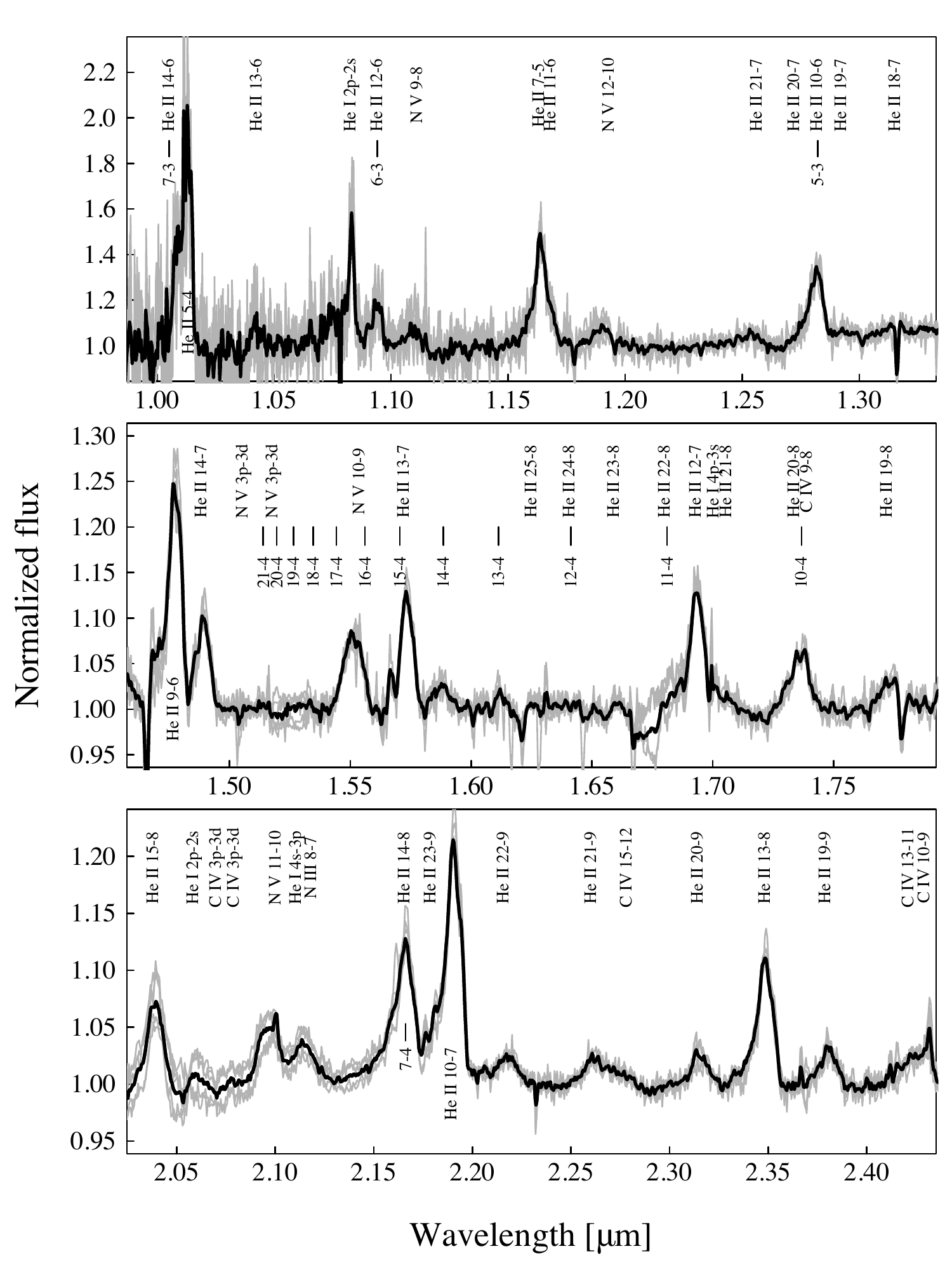}
\end{center}
\vspace{-12pt}
\caption{The 1.0--2.4 $\mu$m, rectified line spectra of Cyg X--3 from X-ray orbital phases $\phi$ = 0.4--0.5 with line identifications. The individual spectra are plotted in grey lines, and their average in black. The tick marks represent the locations of lines from Paschen and Brackett series. The spectrum is shifted to the rest frame of the helium lines.} \label{lines}
\end{figure*} 

To study the line spectrum, we de-reddened the mean spectrum spanning the X-ray orbital phases $\phi$ = 0.4--0.5, which shows the highest equivalent widths of the emission lines, and subtracted a power law from the spectrum. Fig. \ref{lines} shows the resulting observed line spectrum from 1.0--2.4 $\mu$m. The line spectrum is similar to what have been observed in earlier works. It is dominated by ionized helium lines throughout the spectrum. In addition, there are strong, highly-ionized nitrogen lines at 1.55 $\mu$m (N V 10--9) and 2.10 $\mu$m (N V 11--10) as identified already in \citet{fender99}, and we find weaker ones at 1.11 $\mu$m (N V 9--8) and 1.19 $\mu$m (N V 12--10). Possibly, a lower ionization nitrogen line at 2.12 $\mu$m (N III 8--7) is present, but it is likely blended with the He I 4s$^1$S -- 3p$^1$P$^0$ line \citep{vankerkwijk96,fender99}. We identify also ionized carbon lines at 2.43 $\mu$m (a blend of C IV 13--11 and C IV 10--9), 2.28 $\mu$m (C IV 15--12) and possibly at 1.74 $\mu$m (C IV 9--8), but which is blended with He II 20--8 line. As previous $K$-band observations have shown, occasionally the spectrum displays very strong 2.06 $\mu$m He I 2p$^1$P$^0$ -- 2s$^1$S emission line that is present at times when the source is in outburst \citep{vankerkwijk96,fender99}. At other times it shows much weaker emission and/or absorption component, as is the case in our observations. In addition, the 1.08 $\mu$m He I 2p$^3$P$^0$ -- 2s$^3$S line as identified in the $I$-band observations \citep{vankerkwijk96}, is present in our spectra. There are some indications of hydrogen lines from Brackett series in the H-band, but if present their equivalent widths are fairly weak.

\subsection{Line profile} \label{line_profile}

\begin{figure}
\begin{center}
\includegraphics[width=0.35\textwidth]{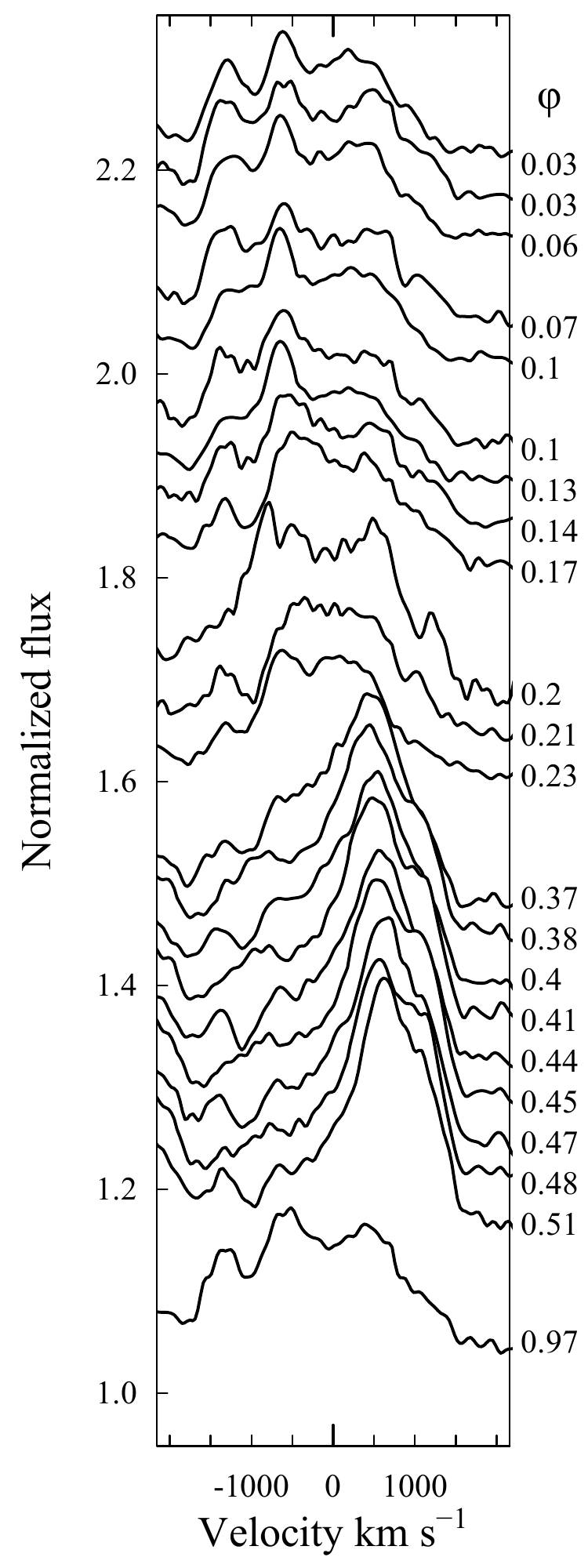}
\end{center}
\vspace{-12pt}
\caption{The line profile of 2.189 $\mu$m He II 10--7 emission line for different X-ray orbital phases ($\phi$). The profiles have been shifted vertically for clarity.} \label{compare_em}
\end{figure}

The line profile changes noticeably with orbital phase, as previously noted by \citet{vankerkwijk96}, \citet{schmutz96} and \citet{fender99}. Fig. \ref{compare_em} shows the line profile of He II 10-7 line at 2.189 $\mu$m with a given X-ray phase (note that the line profiles for different X-ray phases have been taken from different observations). This is the same line that was presented in \citet{schmutz96}, and shows similar line profile with several components. Thus, we can infer that the shape of the line profile is most likely connected to the system geometry and not to any stochastic process, like sudden increases in the wind density in an outburst or a clump in the wind (although clumps probably play a role, as consecutive days show some differences in the line profile). The emission line seems to be composed of several components at different velocities. However, the base of the line is broad and does not seem to change with phase. Most of the lines that are resolved enough present similar structure.      

The most prominent component in the line profile is the red peak around 500 km/s. The equivalent width, and the radial velocity of the red peak changes with the X-ray phase, being strongest/most redshifted at phase $\phi \sim$0.5 and weakest/least redshifted at phase $\phi \sim$0.0. The red peak does not seem to fade away completely in any phase. The blue peak around $-$500 km/s does not become as strong as the red peak, likely arising from an asymmetry in the wind, and/or from constant blue-shifted absorption. There is another peak bluewards around $-$1200 km/s, prominently seen at X-ray phase $\phi$ = 0.1--0.2. Similarly, there seems to be a counterpart on the redshifted side at the same velocity. However, it seems more likely that the blue/red peak are split in two because of absorption at $\sim\pm$1000 km/s.        

\begin{figure}
\begin{center}
\includegraphics[width=0.35\textwidth]{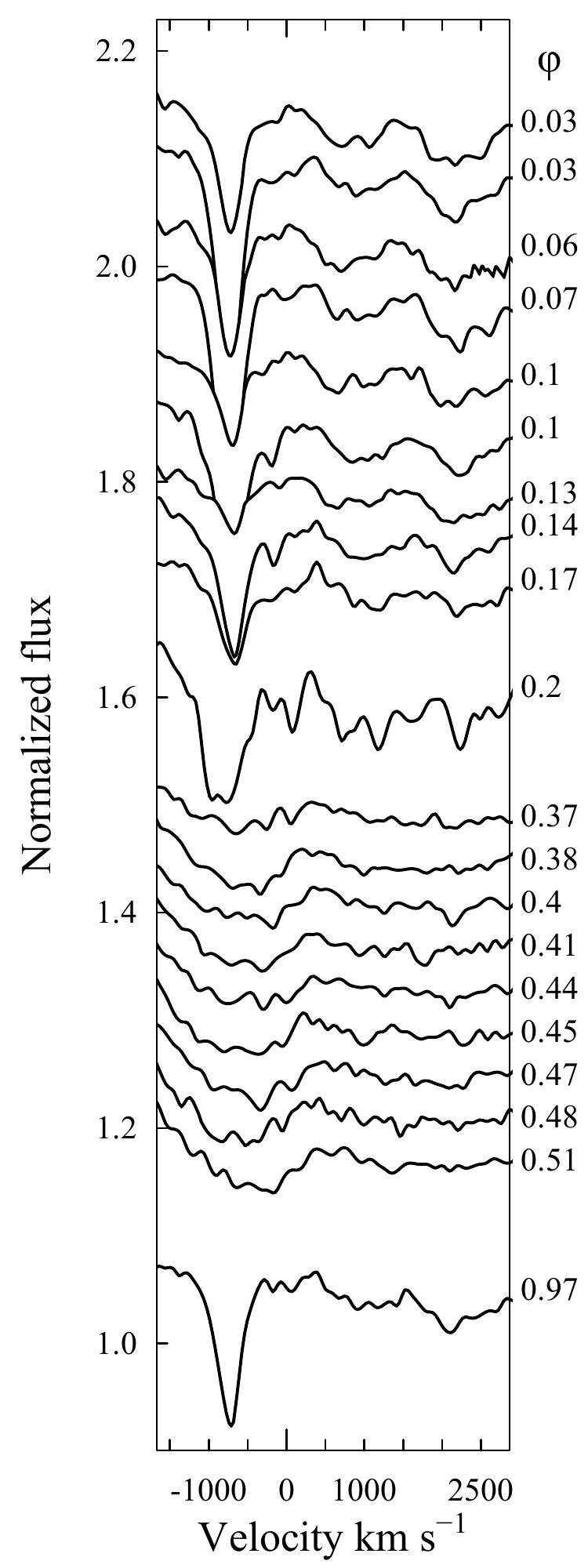}
\end{center}
\vspace{-12pt}
\caption{The line profile of the absorption complex around 2.058 $\mu$m (He I 2p--2s) for different X-ray orbital phases ($\phi$). The profiles have been shifted vertically for clarity.} 
\label{compare_abs}
\end{figure}

The 2.06 $\mu$m absorption line that was used to infer the orbital modulation of the companion star in \citet{hanson00}, could be a part of a more complicated absorption profile. Like in the He II emission line profile, the He I absorption consist of several absorption components at different velocities (Fig. \ref{compare_abs}). The strongest component is blueshifted by $\sim$700 km/s and is prominent and narrow (FWHM $\sim$400 km/s) at X-ray phases around $\phi$ = 0. However, it changes to shallow and wide (FWHM $\sim$1000 km/s) line when the X-ray phase is around $\phi$ = 0.5. Likely, this is because of Hatchett \& McCray effect \citep{hatchett77}, when the X-ray illuminated gas is too ionized to absorb line photons, that are then free to propagate. There are also two redshifted components at velocities $\sim$1000 km/s and $\sim$2250 km/s. Alternatively, these could be blueshifted absorption lines from C IV 3p--3d triplet at 2.071, 2.080 and 2.084 $\mu$m.

Some of the emission lines show P Cygni profiles, most notably He I 2p--2s at 1.08 $\mu$m and He II 9--6. The absorption component of the He I line at 1.08 $\mu$m and 2.06 $\mu$m have been successfully used to determine the terminal velocity \citep{howarth92,eenens94}. We measure the blueshift of the 1.08 $\mu$m He I absorption component as $\sim$1000 km/s. The velocity of the line does not vary much over the orbit. In addition, the blue edge for the 2.06 $\mu$m He I absorption line is $\sim$1000 km/s (see Fig. \ref{compare_abs}, and text above), therefore the terminal velocity is likely lower than 1500 km/s as derived in \citet{vankerkwijk96}, and closer to 1000 km/s.       

\subsection{WR-type and luminosity of the companion star} \label{wrtype} 

Previously, the companion star has been estimated to be a WR star (WN7; \citealt{vankerkwijk96}), based on the near-infrared lines as the optical is heavily absorbed. There are several criteria suggested in the literature for sub-type classification of WR stars. \citet{figer77} and \citet{crowther06} suggest equivalent width ratios of He II 2.189 $\mu$m and He II 2.165 $\mu$m as a good discriminator. The value for the ratio for Cyg X--3 is $\sim$1.6, which corresponds to WR sub-type $<$ WN7. \citet{figer77} consider also the equivalent width ratio of He II 2.189 $\mu$m and He I 4s--3p 2.112 $\mu$m, which is $\sim$4 for Cyg X--3, corresponding to WR sub-types WN 4--6. However, we note that the He I 4s--3p 2.112 $\mu$m line could be blended with the N III 8--7 line and as such can affect to the line ratio estimate. In addition, \citet{crowther06} suggest an additional diagnostic of the equivalent ratio of He II 1.012 $\mu$m and He I 2p--2s 1.083 $\mu$m, which is $\sim$4.3 for Cyg X--3, indicative of a subtype WN 4--5. The above-mentioned equivalent width ratios for Cyg X--3 can be found in Table A2. Furthermore, WN stars are divided to broad-lined or weak-lined stars depending on the FWHM of the emission lines of He II 1.012  $\mu$m and He II 2.189 $\mu$m \citep{crowther06}. The corresponding values for Cyg X--3 range from 40 \AA\/ to 80 \AA\/ for He II 1.012  $\mu$m, and from $\sim$60 \AA\/ to $\sim$110 \AA\/ for He II 2.189 $\mu$m, indicating a weak-lined star. Thus, based on the above criteria, the most likely sub-types for the WR star in Cyg X--3 are WN 4--6. This is also in line with the K-band absolute magnitude estimates of WN4--6 stars \citep[][see Section \ref{spectrum}]{rosslowe15}.

The stellar luminosity for WN 4--6 subtypes is estimated to be log ($L/L_{\odot}$) = 5.2--5.3 \citep{crowther07}. We estimate the luminosity of the WR star as log $L/L_{\odot}$ = $-0.4(M_{v}+BC_{v}-M_{Bol}^{\odot})$, where $M_{Bol}^{\odot}$ is the solar bolometric luminosity taken to be 4.74 \citep{bessell98}, $BC_{v}$ is the bolometric correction estimated to range from $-$4.1 to $-$4.2 for WN6 stars \citep{nugis00,crowther06}, and $M_{v}$ is the narrow-band absolute visual magnitude \citep{smith68} estimated to relate to the absolute K-band magnitude as $M_{v}-M_{K_{S}}=-0.2$ \citep{crowther06}. The narrow-band magnitude is related to the Johnson system as $M_{v}-M_{V}=0.1$ \citep{lundstrom84}. Taking absolute K-band magnitude from Section \ref{spectrum} we derive $L/L_{\odot} = 5.32\pm0.16$ for a distance of 7.4 kpc, and $L/L_{\odot} = 5.59\pm0.12$ for a distance of 10.2 kpc, where the errors include uncertainty in the K-band magnitude and distance. Thus, the luminosity for the closer distance estimate is consistent with the WN 4--6 subtype population.

\subsection{Modeling the spectrum with WR atmosphere models} \label{wrmodel}

\begin{figure*}
\begin{center}
\includegraphics[width=1.0\textwidth]{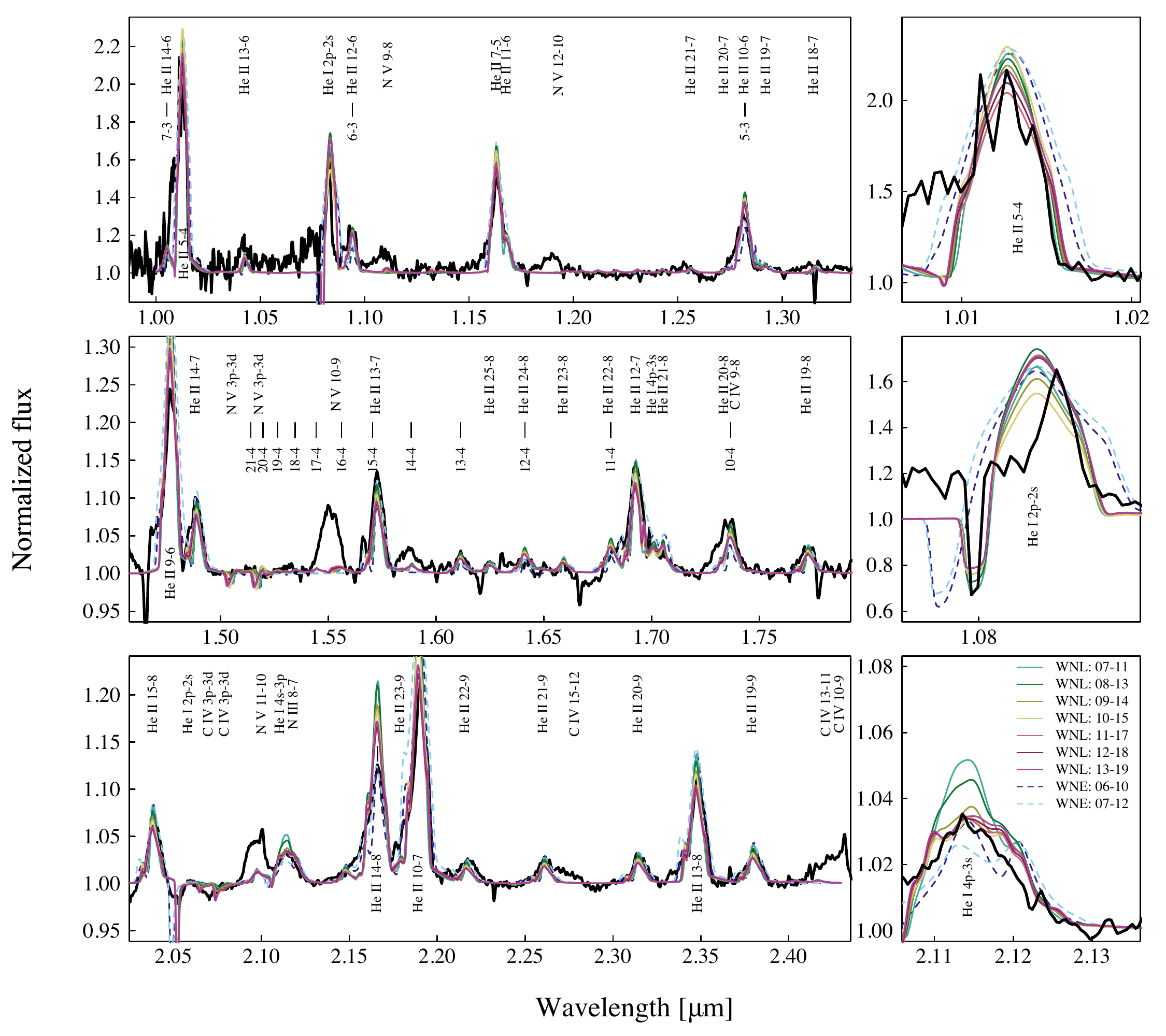}
\end{center}
\vspace{-12pt}
\caption{Left panels show the normalized, average line spectrum from Fig. \ref{lines} (black line), overplotted with the most representative model iterations discussed in the text, and tabulated in Table \ref{params} (coloured lines, solid for WNL grid models and dashed for WNE grid models). The spectrum is shifted to the rest frame of the helium lines. The models differ in transformed radii and are scaled to match the emission line fluxes of the average spectrum to account for the two-temperature wind (see Table \ref{params} and legend in the bottom right panel for the corresponding model identifier). Right panels show three zoomed-in sections from the line spectrum to highlight differences in the line regions of He II 5--4 1.012 $\mu$m (\textit{top}), He I 2p--2s 1.083 $\mu$m (\textit{middle}) and He I 4p--3s 2.112 $\mu$m (\textit{bottom}), important in deriving the ionization balance, and subsequently the parameters of the stellar wind.} \label{compare}
\end{figure*}



To derive further estimates for the stellar parameters we compare the mean observed spectra from X-ray phases $\phi$ = 0.4--0.5 (Fig. \ref{lines}) to the non-LTE atmosphere models of WR stars by \citet{grafener02,hamann03,hamann04,todt15}. Using the WR-type estimate from above, we consider the possibility that the WR star can be an early-type (WN4 to WN6) or a late-type (WN6) star, and therefore we use both the WNE model grid and the WNL model grid (containing 20\% of hydrogen) with the galactic metallicity from the PoWR website\footnote{http://www.astro.physik.uni-potsdam.de/~wrh/PoWR/powrgrid1.php}. The WNE and WNL model grids have similar parameters for the stellar wind with a fixed clumping factor, $D=4$, and luminosity, log $L/L_{\odot}$ = 5.3, that are typical values for WN-type stars \citep{hamann95,hamann98}. The clumping factor of the WR wind in Cyg X--3 has been estimated to be D=3.3--14.3 based on X-ray modeling of the stellar wind \citep{szostek08}. Thus, the model clumping factor is consistent, albeit on the low-side of the probable range. The model grids differ in the terminal velocity which is fixed to $v_{\infty} = 1600$ km/s and $v_{\infty} = 1000$ km/s in the WNE and WNL grids, respectively. Corresponding to the values derived from the He I P Cygni absorption components in Section \ref{line_profile}, the terminal velocity is closer to $v_{\infty} = 1000$ km/s, and thus the WNE model spectra have slightly wider lines and more blueshifted absorption lines, but we do not disregard the WNE grid models based on the higher terminal velocity. 

The grid for the different models is based on two parameters that defines the line spectra: the ``transformed radius'',

\begin{equation}
R_{t} = R_{*} \Bigg[ \frac{v_{\infty}}{2500 \ \mathrm{km \ s}^{-1}} \Bigg/ \frac{\dot{M} \sqrt{\mathrm{D}}}{10^{-4} \ \mathrm{M}_{\odot} \ \mathrm{yr}^{-1}} \Bigg]^{2/3},
\end{equation}

where $R_{*}$ is the ``stellar radius'' (at the radial Rosseland continuum optical depth of 20), and $\dot{M}$ is the mass-loss rate, and the stellar temperature, $T_{*}$, that is connected to the stellar radius and luminosity via Stefan-Boltzmann law $L = 4 \pi \sigma R_{*}^{2} T_{*}^{4}$.  

To find a suitable grid model, we began by searching the grids for those models that produce the same equivalent width ratios of adjacent ionization stages of helium than found in the observations (He II 2.189 $\mu$m / He I 4s--3p 2.112 $\mu$m and He II 1.012 $\mu$m / He I 2p--2s 1.083 $\mu$m, see above and Table A2). This search resulted in those grid models that have effective temperatures of T$_{2/3} \sim$40-50 kK (located at the radial Rosseland continuum optical depth of $2/3$). Note that in the model grid the temperature is defined at the radial Rosseland continuum optical depth of 20, i.e. the stellar temperature, and thus differs from T$_{2/3}$ for denser winds. In addition, the grid models for the very dense winds occur in a degenerate parameter space defined only by the product of the transformed radius and the square of the stellar temperature \citep{hamann04} and thus hotter but smaller WR stars than considered in the model grids can produce similar spectra.

The transformed radius is mostly connected to the line equivalent widths, so that the grid models with the same stellar temperature differ only by their line equivalent widths \citep{schmutz89}. Thus, to find a suitable transformed radius we continue by systematically comparing the mean spectrum to all grid models with T$_{2/3} \sim$40-50 kK in the WNE and WNL grids, but failed to find a representative one to produce the observed He I and He II emission lines. Either the emission lines were too narrow, or over-predicted the lines by a wide margin (see Fig. \ref{spec}). Especially difficult was to reproduce the correct amount of flux in the He I lines at the same time as for the He II lines. We also checked whether relaxing the temperature requirement would result in acceptable model spectra, but for lower temperatures the grid models show strong absorption lines associated with the He II lines and too narrow line profiles as compared to the observed spectra, and for higher temperatures the He I / He II ratio of the grid models was unacceptable. In addition, the models cannot account for the highly-ionized nitrogen and carbon lines observed in the data, but these might arise from the different photoionization regimes as caused by the intense X-ray radiation emitted by the compact object as discussed in Section \ref{windcomp}.

Motivated by the two-temperature model of the stellar wind by \citet{vankerkwijk96}, where most of the line emission takes place in the cool wind behind the WR star that is not irradiated by the intense X-ray emission from the compact object resulting in reduced emission line fluxes as compared to normal WR star atmosphere, we consider modifying the grid model spectra in a simple, qualitative way. Due to the spherical symmetry of the PoWR atmosphere models, we consider that rendering a portion of the atmosphere incapable of forming a line-driven wind by X-ray overionization can be approximated with reducing emission line fluxes by some factor, i.e. the line-driven wind is confined to a cone of some opening angle with the WR star at its vertex. We do not expect the infrared continuum to depend strongly on the X-ray heating, as while the heated part will be less opaque it will emit approximately the same amount of infrared emission as if it would be cool, since its smaller effective emitting area is compensated by its higher temperature. We stress that the line scaling is a very crude approximation of the physical scenario, but necessary to find representative grid model spectra to match the observed spectra. We discuss more about the motivation for the line scaling and possible effects on the value of scaling in Section \ref{caveats}. \citet{vankerkwijk96} estimate that the emission line fluxes are reduced by a factor of a few to account for this effect, and thus we study iteratively whether reducing the emission line fluxes of individual grid models by a given factor will result in a better comparison. 

By scaling the emission lines of the grid models by a factor ranging from 2.5--4.5, we are able to find representative models from both WNE and WNL grids for the mean spectrum shown in Fig. \ref{compare} with correct ratios for the He I and He II lines. The corresponding model parameters of these models are shown in Table \ref{params}. Due to the scaling invariance of the PoWR grid models to different luminosities as long as the transformed radius and stellar temperature are unchanged, we find the correct luminosity by scaling the grid model spectral energy distribution and fitting it to the observation shown in Fig. \ref{spec} (example shown for grid model 7--11 from the WNL grid). The K-band extinction, A$_{K}$, is selected so that the spectral slope of the data matches to that of the model. Subsequently, the stellar radii are scaled with L$^{1/2}$ and the mass-loss rate with L$^{3/4}$ to keep T$^{*}$ and R$_{t}$ unchanged. While the model grids have constant mass for the WR star (12 $\mathrm{M}_{\odot}$), we estimate the mass of the WR using the luminosity-mass relations of \citet{grafener11} (their Eq. 13 for the core He-burning case, since WR stars at solar metallicities most likely do not evolve quasi-homogenously) resulting in a mass range 8--10 $\mathrm{M}_{\odot}$ and 11--14 $\mathrm{M}_{\odot}$ for distances 7.4 kpc and 10.2 kpc, respectively. 

The mass-loss rates of the representative grid models range between a relatively narrow range of $\dot{M} = (0.8-1.1) \times 10^{-5} \ M_{\odot} \ \mathrm{yr}^{-1}$ and $\dot{M} = (1.2--1.8) \times 10^{-5} \ M_{\odot} \ \mathrm{yr}^{-1}$ for the lower and higher distance estimates, respectively. If the clumping factor is taken to be the one derived in \citet{szostek08} (D=3.3--14.3) instead of the model value (D=4), the mass-loss rates are then $\dot{M} = (0.4-1.3) \times 10^{-5} \ M_{\odot} \ \mathrm{yr}^{-1}$ and $\dot{M} = (0.6-2.0) \times 10^{-5} \ M_{\odot} \ \mathrm{yr}^{-1}$.

\begin{table*}
\caption{Stellar parameters of the most representative WR atmosphere models to the Cyg X-3 IR spectra. In the calculations we have used a clumping factor $D = 4$. The columns are: (1) the PoWR grid model identifier, (2) the model atmosphere temperature at $\tau$=20 (stellar temperature), (3) the model transformed radius, (4) the model atmosphere temperature at $\tau$=2/3, (5) the model atmosphere radius at $\tau$=2/3, (6) the model stellar luminosity at distances 7.4/10.2 kpc scaled to match the data, (7) the model stellar radius scaled with the luminosity, (8) the model wind mass-loss rate scaled with the luminosity, (9) the model wind terminal velocity, (10) the mass of the WR star calculated from the luminosity using relations of \citet{grafener11} (the value in the model is fixed to 12 M$_{\odot}$), (11) the scaling factor used to scale the emission line fluxes (12) the spectral slope of the model spectral energy distribution, (13) the K-band extinction needed to match the spectral slope of the data to that of the model, and (14) the hydrogen mass fraction of the model atmospheres.} \label{params}
\begin{center}
\begin{tabular}{lccccccccccccc}
\hline\hline
(1) & (2) & (3) & (4) & (5) & (6) & (7) & (8) & (9) & (10) & (11) & (12) & (13) & (14) \\
Grid & T$_{*}$ & log R$_{t}$ & T$_{2/3}$ & R$_{2/3}$ & log L & R$_{*}$ & log $\dot{M}$ & v$_{\infty}$ & M & f & $\alpha$ & A$_{K}$ & X$_{H}$ \\
model & [kK] & [R$_{\odot}$] & [kK] & [R$_{\odot}$] & [L$_{\odot}$] & [R$_{\odot}$] & [M$_{\odot}$/yr]  & [km/s] & [M$_{\odot}$] & & & & \\
\hline
\textbf{WNL:} \\
07--11 & 50.1 & 1.0 & 45.5 & 7.2 & 5.15/5.43 & 5.0/6.9 & -5.15/-4.94 & 1000 & 10/14 & 3.0 & -3.1 & 1.47 & 0.2 \\ 
08--13 & 56.2 & 0.8 & 44.7 & 7.5 & 5.05/5.34 & 3.5/4.9 & -5.08/-4.86 & 1000 & 9/13 & 3.5 & -3.0 & 1.44 & 0.2 \\ 
09--14 & 63.1 & 0.7 & 47.2 & 6.7 & 5.08/5.36 & 2.9/4.0 & -5.06/-4.85 & 1000 & 9/13 & 4.0 & -2.96 & 1.43 & 0.2 \\ 
10--15 & 70.8 & 0.6 & 49.3 & 6.1 & 5.08/5.36 & 2.3/3.2 & -5.06/-4.84 & 1000 & 9/13 & 4.0 & -2.9 & 1.41 & 0.2 \\  
11--17 & 79.4 & 0.4 & 43.6 & 7.8 & 4.94/5.22 & 1.6/2.2 & -4.99/-4.79 & 1000 & 8/11 & 4.5 & -2.9 & 1.41 & 0.2 \\ 
12--18 & 89.1 & 0.3 & 44.6 & 7.5 & 4.94/5.22 & 1.2/1.7 & -5.03/-4.80 & 1000 & 8/11 & 4.5 & -2.9 & 1.40 & 0.2 \\ 
13--19 & 100 & 0.2 & 44.9 & 7.4 & 4.96/5.24 & 1.0/1.4 & -5.00/-4.78 & 1000 & 8/11 & 4.5 & -2.8 & 1.39 & 0.2 \\ 
\hline
\textbf{WNE:} \\
06--10 & 44.7 & 1.1 & 42.6 & 8.2 & 5.15/5.43 & 6.3/8.7 & -4.95/-4.74 & 1600 & 10/14 & 2.5 & -2.9 & 1.50 & -- \\ 
07--12 & 50.1 & 0.9 & 43.8 & 7.8 & 5.08/5.36 & 3.8/5.3 & -4.98/-4.76 & 1600 & 9/13 & 3.0 & -3.1 & 1.46 & -- \\ 
\hline
\end{tabular}
\end{center}
\end{table*}



\subsection{Radial velocities} \label{rv}

\begin{figure*}
\begin{center}
\includegraphics[width=1.0\textwidth]{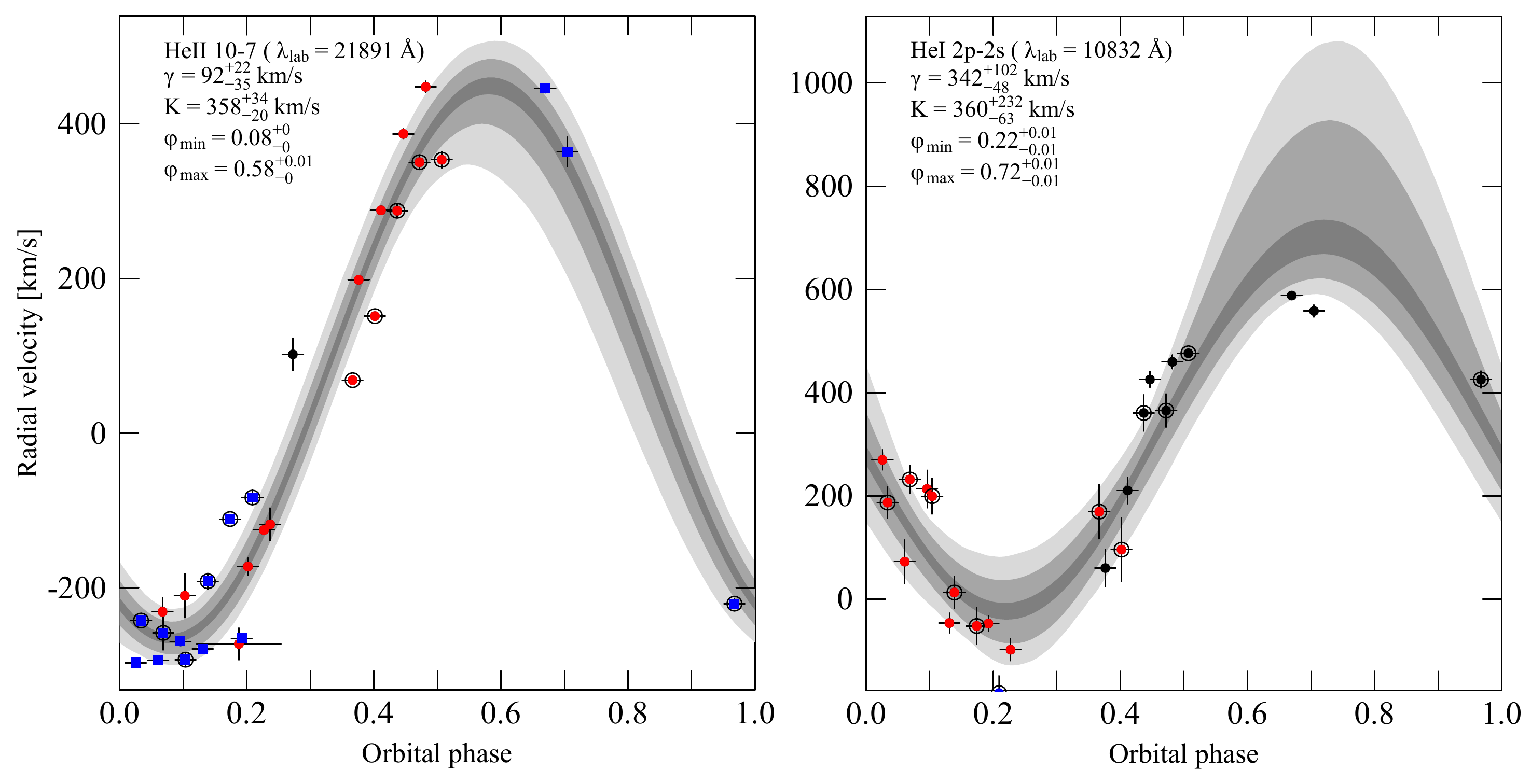}
\end{center}
\vspace{-12pt}
\caption{Radial velocity curves for two emission lines (\textit{left}: He II 10-7, \textit{right}: HeI 2p--2s). Blue squares correspond to spectra where the line is fitted with three gaussian profiles, red points with two gaussian profiles, and black points with one gaussian profile. The final radial velocity is determined as a weighted sum of the gaussian centroids. Encircled points correspond to the longest consecutive set of spectra taken on 16 June 2015. Radial velocity curves are fitted with a sinusoidal function with systemic velocity, radial velocity semi-amplitude, and its minimum and maximum X-ray phase marked in each panel. The grey bands correspond to the 1$\sigma$, 2$\sigma$ and 3$\sigma$ errors on the fit.} \label{rv_em}
\end{figure*}

We measure the systemic velocity, the radial velocity (RV) semi-amplitude and the FWHM of 22 emission lines identified in most of the spectra. These include mostly He II lines (15 different lines), two He I lines (He I 2p--2s and He I 4s--3p, but which could be blended with N III 8--7), three N V lines, and two C IV lines. We fit the line profile with a model consisting of multiple gaussian profiles, depending on the number of peaks. If the line profile is fit with two or more gaussians, the line centroid is defined as the weighted sum of the gaussian fit centroids weighted by their equivalent widths and normalised by the sum of their equivalent widths. After getting the weighted line centroids of all lines, their radial velocity is computed and phase-folded to the X-ray phase. The RV curve is then fitted by a sinusoidal with the corresponding errors calculated by a Monte Carlo methods to obtain values for the systemic velocity, RV semi-amplitude and its minimum and maximum phase. Likewise, the FWHM is calculated from the gaussian fits. If a line profile is fitted with multiple gaussians the FWHM is calculated through the whole profile.

\begin{figure}
\begin{center}
\includegraphics[width=0.5\textwidth]{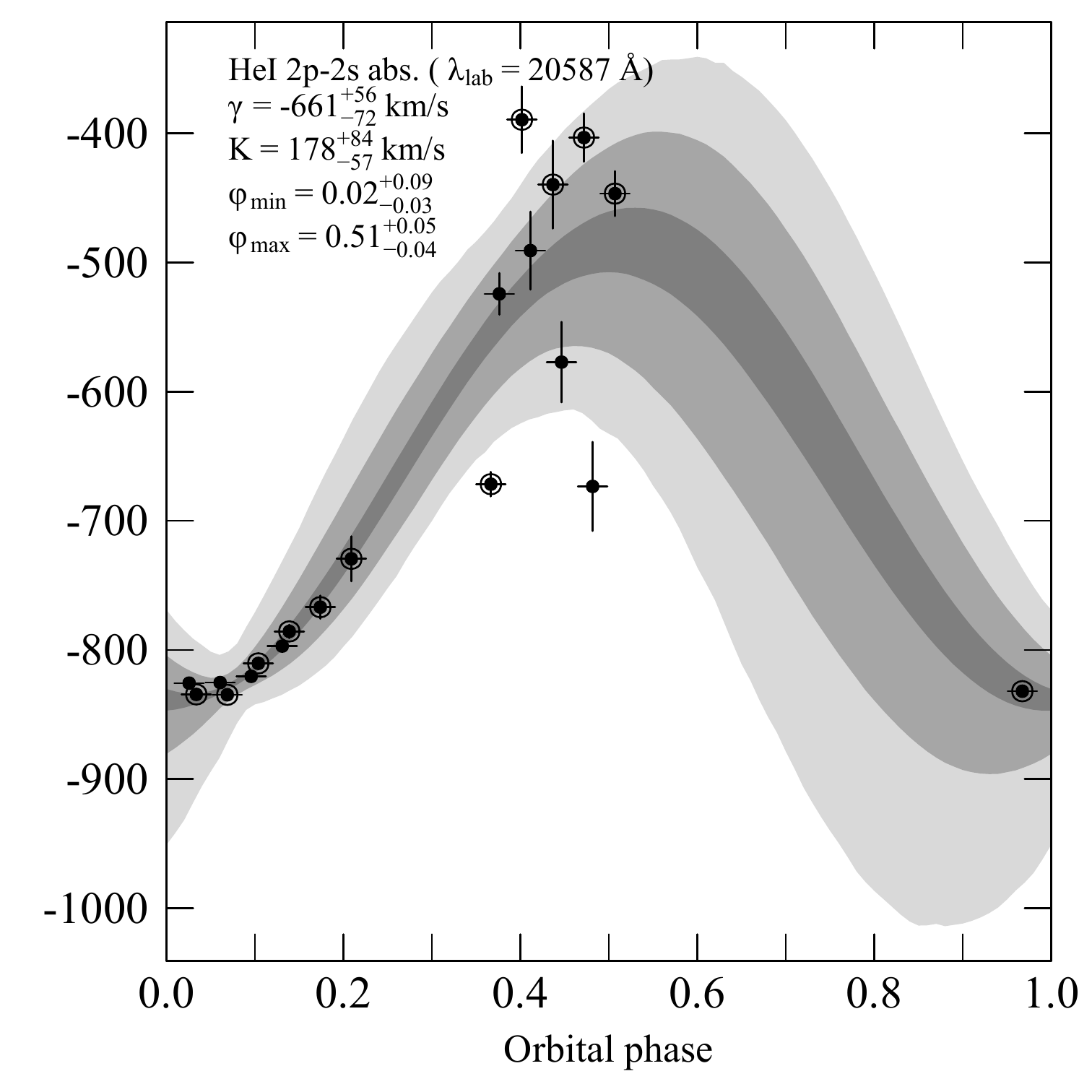}
\end{center}
\vspace{-12pt}
\caption{Radial velocity curve for an absorption line HeI 2p--2s. See the Fig. \ref{rv_em} caption for explanation for the colours and labels.} \label{rv_abs}
\end{figure} 

\begin{figure*}
\begin{center}
\includegraphics[width=0.9\textwidth]{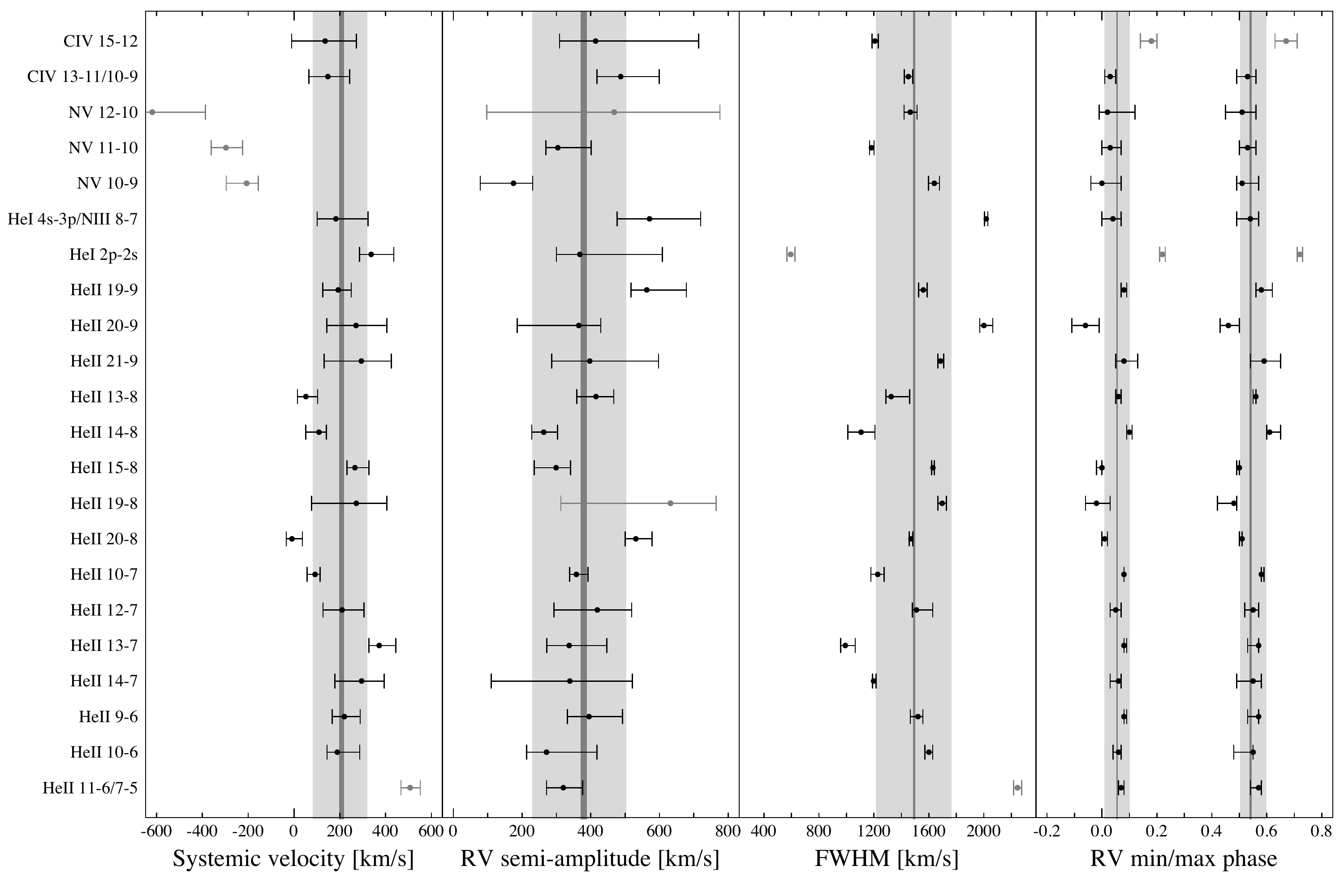}
\end{center}
\vspace{-12pt}
\caption{The systemic velocity (\textit{left}), the radial velocity semi-amplitude (\textit{middle/left}), FWHM (\textit{middle/right}), and the minimum and maximum X-ray phase for the radial velocity semi-amplitude (\textit{right}) of the 22 emission lines identified in most of the spectra. The grey points have not been included in the estimation of the average value (dark grey band) and the 1$\sigma$ error (light grey band).} \label{sys}
\end{figure*}

Despite the spectra being from different dates, the radial velocities remain more or less constant with X-ray phase and the observations taken on different dates follow those taken consecutively (circled points) at 16 June 2015 (see two examples in Fig. \ref{rv_em}). Fig. \ref{sys} shows the collection of systemic velocities, FWHM, RV semi-amplitudes and its minimum and maximum phases of all the emission lines mentioned above. The systemic velocities agree with previous work by \citet{hanson00}: negative for N V and positive for He II emission lines. The mean systemic velocity for the He I / He II / C IV lines is $\gamma = 208^{+113}_{-127}$ km/s (He II 11--6/7--5 is not included in the mean due to it being a blend). The three highly-ionized nitrogen lines, however, have systemic velocity of $\gamma \sim -300$ km/s, completely different from the rest of the lines. The radial velocity semi-amplitudes show some scatter from line to line, but most values cluster around $K = 400$ km/s (the average value being $\bar{K} = 379^{+124}_{-149}$ km/s). This value is consistent with the one derived from X-ray emission lines \citep{vilhu09,zdziarski13} and it points to the infrared radial velocities arising from the orbital modulation of the compact object by shifting the location of the line-forming X-ray shadow of the stellar wind. Almost all lines show similar orbital profile in radial velocity: a RV minimum around phase $\phi$ = 0.05, and a maximum around $\phi$ = 0.55. However, interestingly the He I 2p--2s 1.083 $\mu$m line has a different profile with the minimum and maximum phase shifted by 0.2 (see also Fig. \ref{rv_em}). The FWHM vary quite a lot from line to line,  the average value being FWHM = 1500$\pm$300 km/s. The only exception is again the He I 2p--2s 1.083 $\mu$m line, where the line is narrower with FWHM $\sim$ 600 km/s.  

\section{Discussion} \label{discussion}


\subsection{Caveats of the modeling results} \label{caveats}

In this section, we outline the major caveats of the spectral modeling the infrared spectrum of the WR star in Cyg X-3 with the PoWR atmosphere models, and their impact to the derived parameters. First, the assumption of the spherical atmosphere is most likely not correct. Rather, the wind is flattened to a more equatorial wind \citep{fender99,koljonen13} and/or focused towards the compact object (similar to Cyg X-1). Secondly, the structure and size of the X-ray ionized region is unknown and the effect of X-ray irradiation to the stellar wind is unclear, e.g. the X-ray emission can impair the development of the WR wind. Therefore, the emission line scaling motivated by the simple, two-temperature stellar wind model of \citet{vankerkwijk96} is likely only a rough estimate. In addition, this simplified scenario does not take into account the effect of clumps in the wind that reduces the X-ray ionization \citep{oskinova12}, and impact on the size of the line-forming area. Also, the parameters in the van Kerkwijk model, such as the temperature difference between the hot and cool winds, the system inclination and the opening angle of the cone within which the wind is cool affects to the amount of the wind that is in the cool stage and subsequently to the line profiles. Thirdly, we assume that the line scaling applies similarly to all emission lines, and do not vary between species or ionization stages. Combining all the above physics to one scaling parameter is a crude estimation at best of the situation, but necessary in order to start comparing the WR atmosphere models to the data. The wind inhomogeneities, the degree of X-ray emission and the ionization structure are all important factors and needs to be taken into account in more detailed studies. However, these would need simulating the X-ray ionization structure along the binary orbit together with hydrodynamical modeling and radiation transfer in the stellar wind atmosphere. While these are important issues and merits further studies, these are out of the scope of this paper.   

In addition, we stress that the stellar wind parameters derived from the most representative models (see Table \ref{params}) are an order of magnitude estimates. The mass-loss rate to the compact object can be severely affected, as the ionizing X-ray emission can impair the formation of the stellar wind by weakening or shutting off completely the line-driving mechanism of the X-ray illuminated hemisphere of the WR companion (e.g. \citealt{krticka15,cechura15}). This results in the decrease of the wind velocity and subsequently an increase of the mass accretion rate to the compact object as the Bondi-Hoyle accretion rate is proportional to $v_{\infty}^{-3}$.     

Nevertheless, the mass-loss rates of the grid models are consistent with previous, independent estimates. Based on the stellar wind absorption and emission at soft X-rays, \citet{szostek08} estimated the mass-loss to be $\dot{M} =$ (0.6--1.6) $\times 10^{-5} M_{\odot} \ \mathrm{yr}^{-1}$, where the higher end corresponds to heavier compact object masses ($\sim$20 M$_{\odot}$). Quite similarly, using X-ray modulation curves, \citet{zdziarski12} found $\dot{M} =$ (0.8--1.4) $\times 10^{-5} M_{\odot} \ \mathrm{yr}^{-1}$. \citet{waltman96} placed a limit of $\dot{M} \leq 10^{-5} M_{\odot} \ \mathrm{yr}^{-1}$ from radio flare peak delay times of frequencies 2.25, 8.3 and 15 GHz. However, for stellar wind temperatures exceeding 3.5$\times 10^{4}$ K, which is probably closer to reality based on the models in Section \ref{wrtype}, the limit is raised to $\dot{M} \leq 2.7 \times 10^{-5} M_{\odot} \ \mathrm{yr}^{-1}$. \citet{vankerkwijk93} and \citet{ogley01} estimated the mass-loss rate as $\dot{M} = 4 \times 10^{-5} M_{\odot} \ \mathrm{yr}^{-1}$ and $\dot{M} = (4-30) \times 10^{-5} M_{\odot} \ \mathrm{yr}^{-1}$, respectively, based on the free-free emission from stellar wind \citep{wright75}. However, taking into account the effect of clumping and terminal velocity, these values should be scaled by a factor of 0.17--0.36, bringing the values around $\dot{M} = 10^{-5} M_{\odot} \ \mathrm{yr}^{-1}$. 

\subsection{Stellar wind composition} \label{windcomp}

It is clear, that the nitrogen/carbon fraction is enhanced as compared to the WR spectra of the non-LTE models. The equivalent widths of the N V recombination lines are more in line with an WN2--3 class for the WR companion, however, rest of the line spectra are not. Further clues can be had from the different values of the systemic velocity, that indicates a different location for the line emission in the WR wind. The fact that the systemic velocities have been observed to be approximately the same between observations taken in $\sim$20 years ago (\citealt{hanson00} had $\gamma \sim $150 km/s for He II 10--7 and He II 15--8 and $\gamma \sim -$110 km/s for N V 11--10) suggest that turbulent motion is not likely causing the different systemic velocity of the highly-ionized species. Most likely because of higher ionization energy, N V and C IV survive closer to the compact object and are therefore located in a different region around the system. 



Based on the model spectra, and consistent systemic velocity and radial velocity semi-amplitude values to the He II lines, we can assume that the He I 2p--2s emission line at 1.083 $\mu$m is also produced in the WR wind. The disc contribution is ruled out in \citet{vankerkwijk92} and \citet{fender99}, and they show that the He I lines have to arise from a region much larger than the binary separation. Thus, the quarter phase shift in radial velocity and lower FWHM as compared to other emission lines is an interesting effect that also needs explanation. This might arise again from the ionization structure of the system. Based on hydrodynamical modelling of the wind accretion by X-ray luminous compact object, \citet{kallman15} showed that the ionization structure reflects the structure of the gas density, and that the more dense accretion wake trailing the compact object could provide a location of lower X-ray ionization (their Fig. 12). 



\subsection{Stellar wind structure} \label{structure}

The IR line phenomenology can be understood with a model where the ionization structure of the wind is modified by the intense X-ray emission from the compact object (i.e. the van Kerkwijk model). The emission lines are formed over a large volume that surpasses the size of the orbit (see Table \ref{params} showing that the location of the photosphere, R$_{2/3}$, is further out than the orbital separation approximated to be less than 5 R$_{\odot}$), thus changes in the line profiles are not caused by something happening within the orbit. As the compact object orbits in the WR wind, it photoionizes its surroundings according to its Str\"omgren sphere, thus excavating a region in the wind where the low-ionized emission line formation is suppressed (and line photons cannot be absorbed). In the case where a large fraction of the stellar wind is highly ionized by the X-ray source, the line profile will comprise of two parts: a weak, broad component from the hot, ionized wind, and a strong, narrow component from the cold part of the wind. The latter will move in velocity, as the cold wind region behind the WR star is moving with the compact object and probes different parts of the wind along the orbit with different radial velocities. Indeed, as shown in Section \ref{line_profile}, the width of the base of the emission line seems to stay approximately the same indicating that a weak, broad emission line component from the hot wind region is present at all orbital phases. On top of the hot component a narrower and stronger emission line is imprinted from the cold wind region behind the WR star, changing in radial velocity through the orbit. The redshifted peak is stronger during the inferior conjunction of the compact object (X-ray phase 0.5), as the X-ray ionized wind is incapable of absorbing the line photons, while during superior conjunction (X-ray phase 0.0) the cold wind is partly self-absorbed.

Another possibility is that a shock will form around the compact object as it plows through the WR stellar wind (either due to accretion disc/WR wind, or accretion disc wind/WR wind interface), reminiscent of a wind-wind collision scenario \citep[e.g.][]{stevens99}. The post-shock gas is too hot for the low-ionized emission lines to form, and thus a conical region behind the compact object is excavated. This has the same effect as above; removal of most of the blueshifted emission at X-ray phase 0.5, and likewise for redshifted emission at X-ray phase 0.0. The slight shift of the minimum/maximum RV phase to 0.05/0.55 could be explained by the modulation of the shock cone geometry. In addition, the complicated line profile with several velocity components could arise from the brighter emission in the shock in the heading and trailing side similar to what is observed from wind-wind binaries.



The line profile could be affected by absorption as well. Multiple velocity components seen in the line profile (Figs. \ref{compare_em}, \ref{compare_abs}) can be explained by absorption components at different velocities. On top of the above-mentioned line profile changes due to the X-ray ionization, the red and blue peaks seem to be modified by absorption close to $\pm$1000 km/s. This value coincides with the terminal velocity estimate, and indicates that the line photons are absorbed in the wind. The absorption is stronger around X-ray phase 0.0 (where the blue peak is also stronger) and weakest at X-ray phase 0.5 (where the red peak is stronger). This is most likely due to Hatchett \& McCray effect. Thus, the varying line emission region and the amount of absorption, that are both dependent on the X-ray phase, define the line profile. All the lines seem to be absorbed somewhat similarly. However, there seems to be differences in the amount of absorption in the observations from different days, indicating a modification of the spectra by clumps (see Fig. \ref{compare_em}). As He I 2p--2s 2.058 $\mu$m is not emitted, it presents only the absorption profile. We tracked the RV semi-amplitude of the absorption line of He I 2p--2s 2.058 $\mu$m (Fig. \ref{rv_abs}), and found out that the maximum redshift is occurring at the same X-ray phase as for the He II lines. Thus, we cannot reproduce the results of \citet{hanson00}, where they found that the maximum redshift occurs at X-ray phase 0.2, and cannot assign the He I 2p--2s 2.058 $\mu$m absorption to the WR star. It is more likely that the absorption is taking place in an optically thick region of the stellar wind.  

\subsection{Orbital parameters and the masses of the binary}

Estimating the orbital parameters based on the radial velocities from the IR lines in Cyg X--3 is problematic. Essentially, we are seeing the line forming regions at the distance where the WR star wind becomes optically thin and where they are not ionized by the intense X-ray emission. Thus the velocity distribution reflects primarily that of the stellar wind, rather than that from the motion of the star. Using the same absorption line as in \citet{hanson00}, we could not attribute it to the absorption of the companion star. In addition, \citet{hanson00} used the spectra from an outburst state, when Cyg X--3 displayed very strong and variable He I 2p--2s 2.058 $\mu$m line \citep{fender99}. Thus, the absorption line is located in between the variable, double-peaked structure of the emission line. Therefore, any variations in the peaks of the line, that reflect the ionization structure of the stellar wind, will affect the radial velocity estimate of the absorption line. This places the radial velocity measurement of the WR star in doubt, and this should be verified in the future with detailed IR spectroscopy spanning several orbits of Cyg X--3.

Consequently, this leaves the mass estimates for the binary components uncertain. For the mass of the WR star, we can derive estimates based on theoretical mass-luminosity relations of WR stars by \citet{grafener11}, tabulated in Table \ref{params} and corresponding to a range: $M_{WR}=$ 8--14 $M_{\odot}$, taking into account both distance estimates. \citet{zdziarski13} used the empirical luminosity/mass-loss rate relation of \citet{nugis00} to estimate the mass of the WR star in Cyg X-3. Using this relation (Eq. 6 in \citealt{zdziarski13}), we arrive to a similar range of $M_{WR}=$ 8--15 $M_{\odot}$, assuming that the clumping-corrected mass-loss rate estimate reasonably reflects the value derived in Section \ref{wrmodel}, $\dot{M} = (0.4-2.0) \times 10^{-5} M_{\odot} \ \mathrm{yr}^{-1}$ (but, to which we have mentioned caveats in Section \ref{caveats}).  

Based on the large, positive $P/\dot{P}$ value, and the short orbital period, it is rather unlikely that Cyg X-3 would be a Roche-Lobe overflow (RLOF) system. The system parameters would be quite restricted for this scenario and require small mass for the companion star ($<7 M_{\odot}$; \citealt{lommen05}), that is though borderline with our lower estimate. However, the population synthesis for RLOF WR+BH binaries with such a light companion showed that they traverse the $P/\dot{P}$ value measured for Cyg X-3 in $\sim10^2$ yr, and thus the probability of finding one is negligible \citep{lommen05}. Therefore, it is reasonable to assume that the binary is detached, and accretes through the stellar wind, either by direct capture \citep{ergma98} or through focused wind \citep{friend82}.

This raises concerns about the model stellar radii of the WR star (see Table \ref{params}), that should not exceed the Roche radius (1.3--2.1 $R_{\odot}$ for WR masses 8--14 $M_{\odot}$, and compact object masses 1.4--30 $M_{\odot}$), or as a matter of fact, the orbital separation either ($\sim 3-5 R_{\odot}$). The latter requirement would directly exclude the WNE grid models, with the exception of the grid model 07-12 for a source distance of 7.4 kpc, and one of the WNL grid models. The former would allow only grid models 11-17, 12-18 and 13-19 from the WNL grid. Since these grid models are characterized by very dense winds and lie in the degenerate parameter space defined only by the product of the transformed radius and the square of the stellar temperature \citep{hamann04}, models with higher stellar temperatures but smaller radii are most likely also consistent with the observations. 

\begin{figure}
\begin{center}
\includegraphics[width=0.5\textwidth]{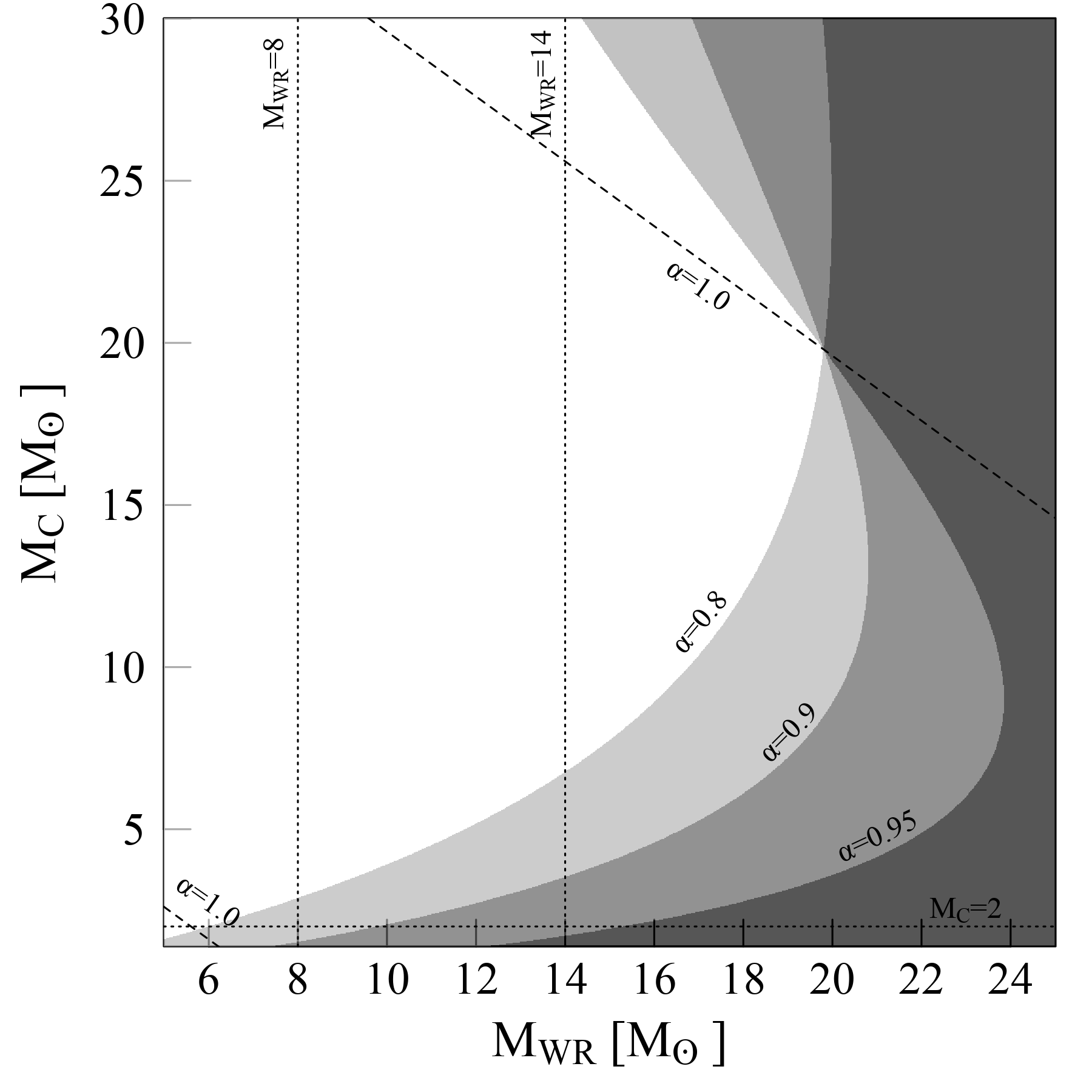}
\end{center}
\vspace{-12pt}
\caption{The allowed masses of the compact object and the WR star in Cyg X-3 using Eqs. \ref{eq2} and \ref{eq3}. The dashed lines delineate a region where $\alpha=1$, i.e. only a tiny fraction ($\lesssim$0.01) of the stellar wind is accreted. The shaded areas show the excluded regions with decreasing values of $\alpha$. The horizontal dotted line mark the maximum mass of a neutron star, and vertical dotted lines delineate the region of the stellar masses derived from the luminosities of the grid models shown in Table \ref{params}.} \label{masses}
\end{figure}
\begin{figure}
\begin{center}
\includegraphics[width=0.5\textwidth]{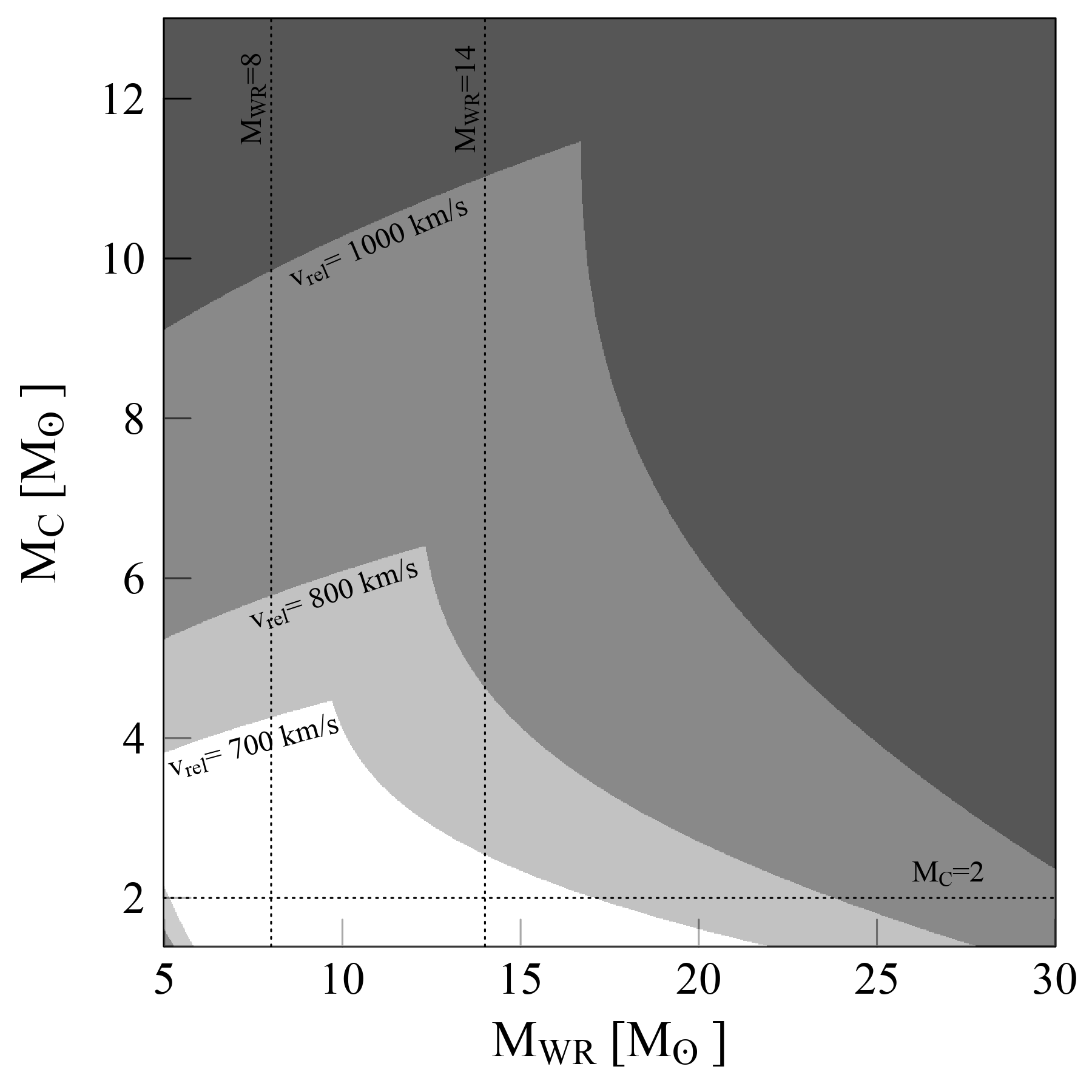}
\end{center}
\vspace{-12pt}
\caption{The allowed masses of the compact object and the WR star in Cyg X-3 using Eq. \ref{eq5} to estimate the amount of wind accreted by the compact object. The shaded areas show the excluded regions with decreasing values of $v_{rel}$.} \label{masses2}
\end{figure}

The total mass of the binary can be estimated from the slowing down of the binary orbit \citep{davidsen74,ergma98,zdziarski13}, with the premise that the binary is detached (see above), and disregarding any effects from tidal interaction (see e.g. \citealt{bagot96}):

\begin{equation} \label{eq2}
M_{WR}+M_{C} \simeq \frac{2\dot{M}P}{\dot{P}} \simeq 190 M_{\odot} \frac{\dot{M}}{10^{-4} M_{\odot} \ \mathrm{yr}^{-1}}, 
\end{equation}

corresponding to $\simeq$ 8 -- 38 M$_{\odot}$ for the above mass-loss rate. This assumes that all the specific angular momentum of the donor is removed by the stellar wind, and only a tiny fraction ($\lesssim$0.01) is accreted onto the compact object. However, if a larger fraction is accreted onto the compact object, e.g. in a focused wind scenario, and re-ejected by forms of accretion wind or jet carrying away the specific orbital angular momentum of the compact object, the situation is more complex and the equation of the period derivative can be expressed as \citep{lommen05}: 

\begin{equation} \label{eq3}
\begin{aligned}
\frac{\dot{P}}{P} = & \frac{\dot{M}}{M_{WR}M_{C}(M_{WR}+M_{C})} \\
& \times \Big[ \Big( 3M_{C}^{2}-2M_{C}M_{WR}-3M_{WR}^{2}\Big)\alpha \\
& - 2M_{C}M_{WR}\beta +3M_{WR}^{2} -3M_{C}^{2}\Big],
\end{aligned}
\end{equation}

where $\alpha$ is the fraction of the mass lost directly from the system via the stellar wind, and $\beta$ is the fraction of the mass accreted by the compact object and re-ejected via accretion wind or jet. In Fig. \ref{masses}, we plot the allowed mass ranges for different values of $\alpha$ and assume that $\beta$ = 1-$\alpha$ (in reality part of the accreted matter can be advected into a black hole, but we find that allowing this alters the results only a tiny fraction), $\dot{P}/P=(1.01-1.05) \times 10^{-6} \ \mathrm{yr}^{-1}$ \citep{zdziarski13} and $\dot{M} = (0.4-2.0) \times 10^{-5} M_{\odot} \ \mathrm{yr}^{-1}$. Estimating $\beta$ is tricky, but an order of magnitude estimate can be had by assuming that the part of an isotropic wind flowing past the accretion radius of the compact object, $r_{acc}$, is accreted:

\begin{equation} \label{eq4}
\dot{M}_{acc} \approx \frac{\pi r^{2}_{acc}}{4\pi a^{2}} \dot{M},
\end{equation}

where $r_{acc} \approx 2GM_{C}v_{rel}^{-2}$, $v_{rel}^{2}=v_{\infty}^{2}+v_{orb}^{2}$, and $a=[P^{2}G(M_{WR}+M_{C})/4\pi^{2}]^{1/3}$. Thus,

\begin{equation} \label{eq5} 
\dot{M}_{acc} \approx 0.0176 \times M_{C}^{2} v_{rel, 1000}^{-4} P^{-4/3}_{4.8} (M_{WR}+M_{C})^{-2/3} \dot{M}, 
\end{equation}

where the masses are in solar units, velocity in units of 1000 km/s and period in units of 4.8 hours. Redefining $\alpha=(1-\beta)$ and $\beta=\dot{M}_{acc}/\dot{M}$, and limiting $\beta<0.25$, corresponding the case where $r_{acc}<a$, we arrive to Fig. \ref{masses2} for the allowed masses with three different, relative wind velocities (1000 km/s, 800 km/s, and 700 km/s) at the location of the compact object. We note that Eq. \ref{eq5} is very sensitive to the relative wind velocity value at the location of the compact object, as can be seen from Fig. \ref{masses2}. As the compact object orbits very close to the WR star, the wind velocity might not have reached the terminal velocity at the location of the compact object, especially if the accretion radius is large, which is directly proportional to the mass of the compact object. Based on the velocity field assumed by the PoWR model atmospheres, the stellar wind velocity reaches 700--800 km/s at the location of the compact object, but could be even lower due to the wind inhibition by the X-ray irradiation as discussed in Section \ref{caveats}. Based on the radial velocity semi-amplitudes of the infrared lines derived in Section \ref{rv} and X-ray lines in \citet{vilhu09,zdziarski13}, the observed orbital velocity is $\sim$400 km/s. Since the orbital solution is unclear, the inclination of the orbit remains uncertain. In any case, it takes most likely a value between 30\degr\/ and 70\degr\/ in order to produce orbital modulation and the lack of orbital dips in the infrared and X-ray lightcurves. Thus, the orbital velocity is then 300--750 km/s, and the relative wind velocity at the location of the compact object 750--1000 km/s. Therefore, the compact object is not very heavy, $< 10 M_{\odot}$, and likely the mass is even lower, $\lesssim 5 M_{\odot}$, for a moderate relative wind velocity. 

\section{Conclusions} \label{conclusions}

In this paper, we have studied the near-infrared spectra of Cyg X--3 during a hard X-ray state. We have confirmed the earlier results that the type of the companion is most likely a WR star. We have found that the 1.0--2.4 $\mu$m spectrum is a representative of a type WN4--6 based on several classification criteria, but the equivalent widths of the emission lines are unusually weak. We were able to produce similar model spectra to the data using the non-LTE atmosphere model of WR stars by \citet{grafener02,hamann03,hamann04,todt15}, assuming that the line forming area is much smaller than the full wind. This scenario was motivated by the van Kerkwijk model \citep{vankerkwijk96}, where the X-ray emission from the compact object ionizes the stellar wind preventing the line formation in the wind, excluding a region behind the WR star where a normal line-driven wind can be launched. However, the models failed to produce the amount of highly-ionized nitrogen and carbon in the spectra, which most likely is due to the complicated ionization structure in the stellar wind. Based on the most representative atmosphere models, we estimate the mass-loss rate to be $\dot{M} = 0.4-2.0 \times 10^{-5} \ M_{\odot} \ \mathrm{yr}^{-1}$, including a reasonable range for the clumping factor, distance, and find that it is consistent with independent estimates derived previously using other wavelengths (radio, X-rays). We have found that scaling the atmosphere models to the observed spectra requires model luminosities log $L_{WR}/L_{\odot}=5.0-5.4$, depending on the distance to the source, that are representative of WN4--6 stars, and that all the infrared flux can be attributed to the WR star with power law indices of the spectral energy distribution ($F_{\lambda} \propto \lambda^{\beta}$) ranging from -2.9 to -3.1 using the K-band extinction $A_{K}=1.4-1.5$. Based on the luminosity-mass relation of \citet{grafener11} this corresponds to masses 8--14 M$_{\odot}$ for the WR star. 

Taking into account the large, positive binary orbital period derivative of Cyg X-3, and the tight 4.8-hour orbit, we have placed limits on the possible mass of the compact object that is less than $\sim 10 M_{\odot}$, and likely even lower, $\lesssim5 M_{\odot}$. However, we cannot rule out a neutron star as a compact object. In addition, as the compact object likely does not accrete via RLOF \citep{lommen05}, the WR star hydrostatic radius cannot exceed the Roche radius, and subsequently places restrictions to the atmosphere models. Thus, it is likely that the WR star in Cyg X-3 is hot ($>$80 kK) with small radius ($<$2 $R_{\odot}$). Its luminosity is $L_{WR}/L_{\odot}=$ 5.0 or 5.2, the mass is 8 or 11 M$_{\odot}$, and the mass-loss rate is $\dot{M} = 0.5-1.1 \times 10^{-5} \ M_{\odot} \ \mathrm{yr}^{-1}$ or $\dot{M} = 0.8-1.8 \times 10^{-5} \ M_{\odot} \ \mathrm{yr}^{-1}$ taking into account a probable range for the clumping factor for distances 7.4 kpc or 10.2 kpc, respectively.



We reinforce the earlier results, that most of the line formation is taking place at the X-ray shadow, outside the Str\"omgren sphere of the ionizing X-ray source. The emission lines are likely formed over a large volume that surpasses the binary orbit. In accordance with the previous results, we find that the systemic velocities of the He II lines and N V lines differ from each other. We argue that this is not due to turbulence, but rather caused by a more stable structure such as a shock or an accretion wake around the compact object that changes the ionization structure in the wind. Detailed ionization structure of the wind, and the locations of the emission line regions could be probed by Doppler tomography using high spectral resolution observations spanning the whole orbit in the future.

Although the spectra displayed the He I 2p--2s 2.058 $\mu$m absorption line that was used in \citet{hanson00} to derive the mass function, we were unable to attribute it to the absorption by the WR star. Rather, the line properties indicate absorption in the wind. In addition, we discussed that the absorption line, as observed during an outburst in the data used by \citet{hanson00}, is located between the highly variable, double-peaked structure of the emission component, that most likely affects the radial velocity estimates. This leaves the published mass function of Cyg X--3 not reliable.

Recently, we have started to tap into the collective sample of WR/black hole binaries, by observing similar sources to Cyg X--3 from other, nearby galaxies, such as in IC 10 and NGC 300. IC 10 X--1 displays similar radial velocity modulation of an He II line with the X-ray phase as the He II lines in Cyg X--3 indicating its origin in the X-ray shadow \citep{laycock15}. Interestingly, in NGC 300 X--1 the radial velocity of an He II line is shifted by 0.5 in phase, possibly indicating highly focussed stellar wind falling in towards the compact object \citep{binder15}. In addition to searching and observing more WR/black hole candidates, detailed observations of the WR stellar wind in Cyg X--3 are important to measure the wind properties along the binary orbit, and subsequently measure the binary properties, in order to compare them with the extragalactic counterparts. WR/black hole binaries are excellent candidates towards forming double black hole binaries, and their compact object mass distribution is important in determining the evolution of massive stars. The fate of Cyg X--3 has been calculated to be a double black hole binary if the WR mass is 14.2 $M_{\odot}$, and the compact object mass 4.5 $M_{\odot}$ \citep{belczynski13}, and we can lend support to this scenario based on our mass estimates for the binary components. In addition, the WR/black hole binaries could be candidates for the lower luminosity ultra-luminous X-ray binaries radiating at $10^{38}-10^{39}$ erg/s.    

\section*{Acknowledgements}

We would like to thank the referees for their recommendations that have improved this paper. We thank Saeqa Vrtilek and Michael McCollough for insightful comments. This work is based on observations obtained at the Gemini Observatory, which is operated by the Association of Universities for Research in Astronomy, Inc., under a cooperative agreement with the NSF on behalf of the Gemini partnership: the National Science Foundation (United States), the National Research Council (Canada), CONICYT (Chile), Ministerio de Ciencia, Tecnolog\'{i}a e Innovaci\'{o}n Productiva (Argentina), and Minist\'{e}rio da Ci\^{e}ncia, Tecnologia e Inova\c{c}\~{a}o (Brazil).

\bibliographystyle{mnras}

\bibliography{references}

\appendix

\section[]{Observation log and line ratios}

\begin{table*}
\caption{Observation log} \label{obs}
\begin{center}
\begin{tabular}{lcccccccc}
\hline\hline
Spectrum & Date & UT start & UT stop & Phase & Phase & Phase \\ 
N$^{\circ}$ & dd/mm/yy & hh:mm:ss.s & hh:mm:ss.s & start & stop & average \\
\hline
1 & 13/06/15 & 13:57:25.1 & 14:07:24.1 & 0.359 & 0.393 & 0.38 \\
2 & 13/06/15 & 14:07:31.6 & 14:17:31.1 & 0.395 & 0.428 & 0.41 \\
3 & 13/06/15 & 14:17:38.1 & 14:27:37.6 & 0.430 & 0.463 & 0.45 \\
4 & 13/06/15 & 14:27:44.6 & 14:37:44.1 & 0.465 & 0.498 & 0.48 \\
5 & 16/06/15 & 11:53:15.1 & 12:03:12.6 & 0.951 & 0.984 & 0.97 \\
6 & 16/06/15 & 12:12:18.6 & 12:22:16.1 & 0.017 & 0.050 & 0.03 \\
7 & 16/06/15 & 12:22:23.1 & 12:32:21.1 & 0.052 & 0.085 & 0.07 \\
8 & 16/06/15 & 12:32:28.1 & 12:42:26.1 & 0.087 & 0.120 & 0.10 \\
9 & 16/06/15 & 12:42:33.1 & 12:52:31.1 & 0.122 & 0.155 & 0.14 \\
10 & 16/06/15 & 12:52:38.1 & 13:02:36.1 & 0.157 & 0.191 & 0.17 \\
11 & 16/06/15 & 13:02:43.1 & 13:12:41.1 & 0.192 & 0.226 & 0.21 \\
12 & 16/06/15 & 13:48:04.1 & 13:58:02.1 & 0.350 & 0.383 & 0.37 \\
13 & 16/06/15 & 13:58:09.1 & 14:08:07.1 & 0.385 & 0.418 & 0.40 \\
14 & 16/06/15 & 14:08:14.1 & 14:18:12.1 & 0.420 & 0.453 & 0.44 \\
15 & 16/06/15 & 14:18:19.1 & 14:28:17.1 & 0.455 & 0.488 & 0.47 \\
16 & 16/06/15 & 14:28:24.1 & 14:38:22.1 & 0.490 & 0.524 & 0.51 \\
17 & 17/06/15 & 12:07:46.8 & 12:17:21.3 & 0.009 & 0.042 & 0.03 \\ 
18 & 17/06/15 & 12:17:51.8 & 12:27:26.8 & 0.044 & 0.077 & 0.06 \\
19 & 17/06/15 & 12:27:57.3 & 12:37:31.8 & 0.079 & 0.112 & 0.10 \\
20 & 17/06/15 & 12:38:02.3 & 12:47:37.3 & 0.114 & 0.147 & 0.13 \\
21 & 17/06/15 & 12:55:40.3 & 13:05:14.8 & 0.175 & 0.209 & 0.19 \\
22 & 17/06/15 & 13:05:45.3 & 13:15:20.3 & 0.211 & 0.244 & 0.23 \\
23 & 06/07/15 & 14:31:02.4 & 14:40:41.0 & 0.653 & 0.686 & 0.67 \\
24 & 06/07/15 & 14:41:11.0 & 14:50:41.0 & 0.688 & 0.721 & 0.70 \\
25 & 29/07/15 & 06:38:16.8 & 06:47:52.8 & 0.185 & 0.219 & 0.20 \\
26 & 29/07/15 & 06:48:22.8 & 06:57:52.8 & 0.220 & 0.253 & 0.24 \\
27 & 05/11/15 & 07:08:52.8 & 07:18:50.8 & 0.051 & 0.084 & 0.07 \\
28 & 05/11/15 & 07:18:58.3 & 07:28:56.3 & 0.086 & 0.119 & 0.10 \\
29 & 05/11/15 & 07:29:03.3 & 07:39:01.3 & 0.121 & 0.254 & 0.19 \\
30 & 05/11/15 & 08:07:53.8 & 08:17:52.8 & 0.256 & 0.289 & 0.27 \\
\end{tabular}
\end{center}
\end{table*}

\begin{table*}
\caption{Equivalent width ratios. The bottom row gives a sample mean and standard deviation.} \label{lineprop}
\begin{center}
\begin{tabular}{lccccccccccccccccccccccc}
\hline\hline
Spectrum & He II 2.189 $\mu$m / & He II 2.189 $\mu$m / & He II 1.012 $\mu$m / \\
N$^{\circ}$ & He II 2.165 $\mu$m & He I 4s--3p 2.112 $\mu$m & He I 2p--2s 1.083 $\mu$m \\
\hline
1 & 1.72$\pm$0.01 & 5.1$\pm$0.2 & -- \\
2 & 1.61$\pm$0.01 & 4.4$\pm$0.1 & 5.9$\pm$0.6 \\
3 & 1.61$\pm$0.01 & 5.3$\pm$0.2 & -- \\ 
4 & 1.47$\pm$0.01 & 5.4$\pm$0.2 & -- \\
5 & 1.57$\pm$0.01 & 2.6$\pm$0.1 & 3.9$\pm$0.4 \\
6 & 1.53$\pm$0.01 & -- & -- \\
7 & 2.2$\pm$0.1 & 2.9$\pm$0.1 & -- \\
8 & 1.39$\pm$0.01 & 3.0$\pm$0.1 & -- \\
9 & 1.47$\pm$0.02 & 3.3$\pm$0.1 & -- \\
10 & 1.60$\pm$0.01 & 2.9$\pm$ & 4.0$\pm$0.4 \\
11 & 1.59$\pm$0.02 & 3.6$\pm$0.1 & 4.0$\pm$0.6 \\ 
12 & 1.71$\pm$0.01 & 5.2$\pm$0.1 & 3.4$\pm$0.5 \\ 
13 & 1.61$\pm$0.01 & 5.0$\pm$0.1 & 4.7$\pm$0.6 \\
14 & 1.48$\pm$0.01 & 4.0$\pm$0.1 & -- \\
15 & 1.53$\pm$0.01 & 5.0$\pm$0.2 & -- \\
16 & 1.69$\pm$0.01 & 4.5$\pm$0.1 & -- \\
17 & 1.49$\pm$0.01 & 2.7$\pm$0.1 & -- \\
18 & 1.72$\pm$0.02 & 2.7$\pm$0.1 & 3.8$\pm$0.4 \\
19 & 1.52$\pm$0.01 & 3.5$\pm$0.1 & 4.5$\pm$0.3 \\
20 & 1.70$\pm$0.01 & 3.3$\pm$0.1 & 4.4$\pm$0.4 \\
21 & 1.54$\pm$0.01 & 3.6$\pm$0.1 & 4.1$\pm$0.3 \\
22 & 1.63$\pm$0.01 & 3.0$\pm$0.1 & 4.9$\pm$0.3 \\
23 & 1.95$\pm$0.02 & -- & -- \\
24 & 2.0$\pm$0.1 & -- & -- \\
25 & 1.62$\pm$0.03 & 4.3$\pm$0.2 & -- \\
26 & 1.52$\pm$0.06 & 4.6$\pm$0.3 & -- \\
27 & 1.33$\pm$0.05 & 4.7$\pm$0.4 & -- \\
28 & 1.24$\pm$0.06 & 2.8$\pm$0.3 & -- \\
29 & 1.41$\pm$0.06 & 2.3$\pm$0.2 & -- \\
30 & 1.3$\pm$0.1 & 2.4$\pm$0.4 & -- \\
\hline
& \textbf{1.6$\pm$0.2} & \textbf{4$\pm$1} & \textbf{4.3$\pm$0.7} \\
\hline

\end{tabular}
\end{center}
\end{table*} 

\label{lastpage}

\end{document}